\documentclass[12pt,draftclsnofoot, perreview, onecolumn]{IEEEtran}
\usepackage{amsfonts}
\usepackage{amsmath}
\usepackage{amssymb}
\usepackage{booktabs}
\usepackage{balance}
\usepackage{color}
\usepackage{citesort}
\usepackage{graphicx}
\usepackage{mathrsfs}
\usepackage{mdwmath}
\usepackage{multirow}
\usepackage{setspace}
\usepackage{subfigure}
\normalsize

\ifCLASSINFOpdf
\else
\fi

\newcommand{\Layerk}{K}
\newcommand{\Allk}{k}
\newcommand{\Alln}{n}

\newcommand{\Memory}{m}

\newcommand{\Qfun}[1]{Q\left(#1\right)}

\ifCLASSOPTIONonecolumn
\newcommand{\figwidth}{0.67\textwidth}

\fi
\ifCLASSOPTIONtwocolumn
\newcommand{\figwidth}{0.48\textwidth}

\fi

\newtheorem{theorem}{Theorem}

\newtheorem{corollary}{Corollary}

\newtheorem{algorithm}{\textbf{Algorithm}}
\newtheorem{example}{\textbf{Example}}
\makeatletter
    
    \newcommand{\Rmnum}[1]{\expandafter\@slowromancap\romannumeral #1@}
\makeatother

\begin{document}
\title{Systematic Block Markov Superposition Transmission of Repetition Codes}
%

\author{Xiao~Ma,~\IEEEmembership{Member,~IEEE}, Kechao~Huang,~\IEEEmembership{Student~Member,~IEEE},
        and~Baoming~Bai,~\IEEEmembership{Member,~IEEE}
        \thanks{This work was partially supported by the $973$ Program (No. $2012$CB$316100$), the $863$ Program (No. $2015$AA$01$A$709$), and the China NSF (No. 91438101 and No. 61172082).}
        \thanks{X.~Ma and K.~Huang are with the School of Data and Computer Science, Sun Yat-sen University, Guangzhou 510006, China~(e-mail:~maxiao@mail.sysu.edu.cn; hkech@mail2.sysu.edu.cn).}
        \thanks{B.~Bai is with the State Key Laboratory of Integrated Service Networks, Xidian University, Xi'an 710071, China (e-mail:~bmbai@mail.xidian.edu.cn).}
}



\maketitle
\IEEEpeerreviewmaketitle

\begin{abstract}
In this paper, we propose systematic block Markov superposition transmission of repetition~(BMST-R) codes, which can support a wide range of code rates but maintain essentially the same encoding/decoding hardware structure. The systematic BMST-R codes resemble the classical rate-compatible punctured convolutional~(RCPC) codes, except that they are typically non-decodable by the Viterbi algorithm due to the huge constraint length induced by the block-oriented encoding process. The information sequence is partitioned equally into blocks and transmitted directly, while their replicas are interleaved and transmitted in a block Markov superposition manner. By taking into account that the codes are systematic, we derive both upper and lower bounds on the bit-error-rate~(BER) under maximum {\em a posteriori}~(MAP) decoding. The derived lower bound reveals connections among BER, encoding memory and code rate, which provides a way to design good systematic BMST-R codes and also allows us to make trade-offs among efficiency, performance and complexity. Numerical results show that:~1)~the proposed bounds are tight in the high signal-to-noise ratio~(SNR) region;~2)~systematic BMST-R codes perform well in a wide range of code rates; and~3)~systematic BMST-R codes outperform spatially coupled low-density parity-check~(SC-LDPC) codes under an equal decoding latency constraint.
\end{abstract}


\section{Introduction}
Since the invention of turbo codes~\cite{Berrou93} and the rediscovery of low-density parity-check~(LDPC) codes~\cite{Gallager63}, constructing practical good codes has been being an active research topic in our field. Recent developments include the invention of polar codes~\cite{Arikan09} and flourishment of spatially coupled LDPC~(SC-LDPC) codes~(first introduced as LDPC convolutional codes~\cite{Felstrom99} and later recast as SC-LDPC codes~\cite{Kudekar11}), both of which are provable capacity-achieving~\cite{Arikan09,Lentmaier10,Kudekar11,Kudekar13} over memoryless binary-input symmetric-output channels. Despite this success in theory, more flexible constructions are still desired in practice. Especially, it is often desirable in practice to design codes that support a variety of code rates but maintain essentially the same encoding/decoding hardware structure. One way to achieve this is the use of rate-compatible codes, which can be constructed from a mother code by using the puncturing and/or extending techniques. The former starts with a low-rate mother code and punctures some coded bits to achieve higher rates~\cite{Hagenauer88,Acikel99,Rowitch00,Ha04,Nik07}, while the latter starts with a high-rate code and extends its parity-check matrix to achieve lower rates~\cite{Yazdani04,Khamy09,NguyenTV12,NguyenTV13,Chen15}. Both puncturing and extending require optimizations. For example, the puncturing patterns for rate-compatible punctured convolutional~(RCPC) codes in~\cite{Hagenauer88} were selected by maximizing the average free distance, while the puncturing distributions for rate-compatible LDPC codes in~\cite{Yazdani04} were optimized by density evolution. In~\cite{Chen15}, the incremental protomatrices for protograph-based raptor-like~(PBRL) LDPC codes were chosen by maximizing the density evolution threshold. To reduce the construction complexity caused by the optimizations, one can use random puncturing, as proposed in~\cite{Mitchell_16JSAC}. However, similar to the conventional punctured LDPC codes~\cite{Yazdani04}, the performance of the randomly punctured LDPC codes degrades significantly when the puncturing fraction increases beyond a threshold. To the best of our knowledge, no methods were reported along with simulations in the literature that can construct good rate-compatible codes over all rates of interest in the interval (0,1).


Recently, a coding scheme called block Markov superposition transmission~(BMST) of short codes~(referred to as \emph{basic codes}) was proposed~\cite{Ma15}, which has a good performance over the binary-input additive white Gaussian noise~(AWGN) channel. It has been pointed out in~\cite{Ma15} that any short code~(linear or nonlinear) with fast encoding algorithm and efficient soft-in soft-out~(SISO) decoding algorithm can be chosen as the basic code. A BMST code is indeed a convolutional code with extremely large constraint length, which has a simple encoding algorithm and a low complexity sliding window decoding algorithm. More importantly, BMST codes have near-capacity performance~(observed by simulation and confirmed by extrinsic information transfer~(EXIT) chart analysis~\cite{Huang15JSAC}) in the waterfall region of the bit-error-rate (BER) curve and an error floor~(predicted by analysis) that can be controlled by the encoding memory. In~\cite{Liang15}, short Hadamard transform~(HT) codes are taken as the basic codes, resulting in a class of multiple-rate codes with fixed code length, referred to as BMST-HT codes. An even simpler construction for multiple-rate BMST codes was proposed in~\cite{Hu14a}, where the involved basic codes consist of repetition~(R) codes and single-parity-check~(SPC) codes, resulting in BMST-RSPC codes. Different from BMST-HT codes which adjust their code rates by setting properly the number of frozen bits in the short HT codes, BMST-RSPC codes adjust the code rates by time-sharing between the R code and the SPC code. The construction of BMST codes is flexible, in the sense that it applies to all code rates of interest in the interval (0,1). However, original BMST codes~\cite{Ma15,Huang15JSAC,Liang15,Hu14a} are neither rate-compatible nor systematic. Note that systematic codes may be more attractive in practical applications since the information bits can be extracted directly from the estimated codeword. Even worse, original BMST codes do not perform well over block fading channels due to errors propagating to successive decoding windows.

In this paper, we propose systematic BMST of repetition codes, referred to as systematic BMST-R codes. For encoding, the information sequence is partitioned equally into blocks and transmitted directly, while their replicas are interleaved and transmitted in a block Markov superposition manner. For decoding, a sliding window decoding algorithm with a tunable decoding delay can be implemented, as with SC-LDPC codes~\cite{Lentmaier10,Iyengar12}. Systematic BMST-R codes not only preserve the advantages of the original non-systematic BMST codes, namely, low encoding complexity, effective sliding window decoding algorithm and predictable error floors, but also have improved decoding performance especially in short-to-moderate decoding latency.

The main contributions of this paper include:
\begin{enumerate}
  \item We propose systematic rate-compatible BMST-R codes by using both extending and puncturing. The construction requires no optimization but applies universally to all code rates varying ``continuously" from zero to one.
  \item We propose an upper bound on the BER of a systematic BMST-R code ensemble under maximum {\em a posteriori}~(MAP) decoding, which can be evaluated by calculating partial input-redundancy weight enumerating function~(IRWEF) with truncated information weight.
  \item We propose a lower bound on the BER of a systematic BMST-R code ensemble under MAP decoding, which depends on the encoding memory and code rate. The derived lower bound reveals connections among BER, encoding memory and code rate, which provides a way to design good systematic BMST-R codes and also allows us to make trade-offs among efficiency, performance and complexity.
  \item We investigate the impact of various parameters on the performance of systematic BMST-R codes, and then present a performance comparison of systematic BMST-R codes and SC-LDPC codes on the basis of equal decoding latency.
\end{enumerate}

Simulation results show that: 1)~the upper and lower bounds are tight in the high signal-to-noise ratio~(SNR) region; 2)~with a moderate decoding delay, the BER curves can match the respective lower bounds in the low BER region, implying that the iterative sliding window decoding algorithm is near optimal; 3)~systematic BMST-R codes perform well~(within one dB away from the corresponding Shannon limits) in a wide range of code rates, confirming the effectiveness of the construction procedure; and 4)~over both AWGN channels and block fading channels, systematic BMST-R codes, overcoming the weakness of non-systematic BMST codes, can have better performance than SC-LDPC codes in the waterfall region under the equal decoding latency constraint.

The rest of the paper is structured as follows. In Section~\ref{SecII}, we present the encoding and decoding algorithms of systematic BMST-R codes. In Section~\ref{SecIII}, we analyze the performance and complexity of systematic BMST-R codes. Numerical analysis and performance comparison are presented in Section~\ref{SecIV}. Finally, some concluding remarks are given in Section~\ref{sec:Conclusion}.

\section{Systematic BMST-R Codes}\label{SecII}
\subsection{Encoding Algorithm}\label{subsec:encoding}
\begin{figure}[t]
   \centering
   \includegraphics[angle=270, clip, width=\figwidth]{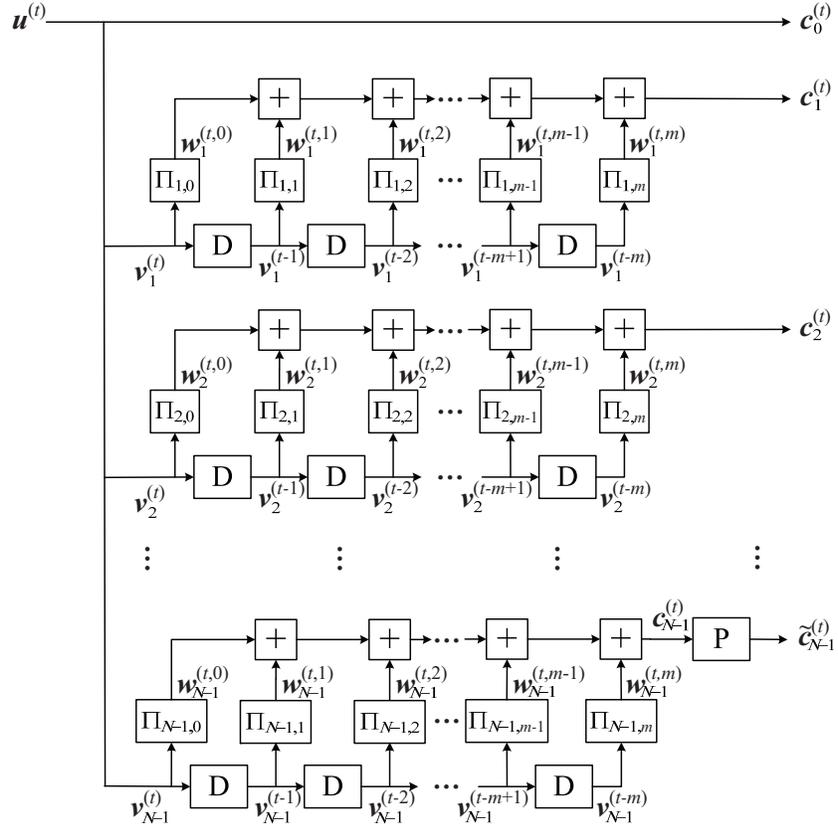}
   \caption{Encoder of a systematic BMST-R code with repetition degree $N$ and encoding memory $\Memory$, where the information subsequence $\boldsymbol{u}^{(t)}$ at time $t$ is encoded into the subcodeword $\boldsymbol{c}^{(t)}=\{\boldsymbol{c}_0^{(t)},\boldsymbol{c}_1^{(t)},\boldsymbol{c}_2^{(t)},\cdots,\widetilde{\boldsymbol{c}}_{N-1}^{(t)}\}$ for transmission.}
   \label{BMST_encoder}
\end{figure}

Let $\mathbb{F}_2 = \{0, 1\}$ be the binary field. Let $\boldsymbol{u}=(\boldsymbol{u}^{(0)}$, $\boldsymbol{u}^{(1)}$, $\cdots)$ be the information sequence to be transmitted, where $\boldsymbol{u}^{(t)} \in \mathbb{F}_2^\Layerk$ is the information subsequence of length $\Layerk$. The encoding algorithm of a systematic BMST-R code of rate $1/N$ with encoding memory $m$ is described as follows~(see Fig.~\ref{BMST_encoder} for reference), where $\boldsymbol{\Pi}_{i,j}$ $(1 \leq i \leq N-1,~0 \leq j \leq \Memory)$ are interleavers of size $\Layerk$.

\vspace{0.10cm}
\begin{algorithm}{Encoding of Systematic BMST-R Codes}\label{alg:encoding}
\begin{enumerate}
  \item{\bf{Initialization}:} \label{step:encoding_initialize} For $t < 0$ and $1 \leq i \leq N-1$, set $\boldsymbol{v}_i^{(t)} = \boldsymbol{0} \in \mathbb{F}_2^\Layerk$.
  \item{\bf{Loop}:} \label{step:encoding_iteration} For $t \geq 0$,
        \begin{itemize}
          \item Repeat $\boldsymbol{u}^{(t)}$ $N$ times such that $\boldsymbol{c}_0^{(t)}= \boldsymbol{u}^{(t)} \in \mathbb{F}_2^\Layerk$ and $\boldsymbol{v}_i^{(t)}= \boldsymbol{u}^{(t)} \in \mathbb{F}_2^\Layerk$ for $1 \leq i \leq N-1$;
          \item For $1 \leq i \leq N-1$,
              \begin{enumerate}
                \item For $0\leq j \leq m$, interleave $\boldsymbol{v}_i^{(t-j)}$ into $\boldsymbol{w}_i^{(t,j)}$ using the $(i,j)$-th interleaver $\boldsymbol{\Pi}_{i,j}$;
                \item Compute $\boldsymbol{c}_i^{(t)} = \sum_{0\leq j \leq m} \boldsymbol{w}_i^{(t,j)}$.
              \end{enumerate}
          \item Take $\boldsymbol{c}^{(t)}=\{\boldsymbol{c}_0^{(t)},\boldsymbol{c}_1^{(t)},\boldsymbol{c}_2^{(t)},\cdots,\boldsymbol{c}_{N-1}^{(t)}\}$ as the $t$-th block of transmission.
        \end{itemize}
  \end{enumerate}
\end{algorithm}
\vspace{0.10cm}

The above encoding structure can implement all code rates of the form $1/N$, $N = 2, 3, \cdots$. If $\Layerk_p$ of $\Layerk$ bits in $\boldsymbol{c}_{N-1}^{(t)}$ are randomly punctured resulting in $\widetilde{\boldsymbol{c}}_{N-1}^{(t)}$, we can implement a code rate $\frac{1}{N-\theta} \in (\frac{1}{N}, \frac{1}{N-1})$, where $\theta \stackrel{\Delta}{=} \frac{\Layerk_p}{\Layerk}$ is the puncturing fraction. In practice, the code need to be terminated. This can be done easily by driving the encoder to the zero state with a zero-tail of length $m\Layerk$ after $L$ blocks of data. That is, for $t = L$, $L+1$, $\cdots$, $L+m-1$, we set $ \boldsymbol{u}^{(t)} = \boldsymbol{0} \in \mathbb{F}_2^\Layerk$, compute $\boldsymbol{c}^{(t)}$ following~{\bf Loop} in Algorithm~\ref{alg:encoding}, and then take the redundant check part of $\boldsymbol{c}^{(t)}$ as the $t$-th block of transmission. The rate of the resulting terminated systematic BMST-R code is
\begin{eqnarray}\label{BMST_RateL}
  R_L &=& \frac{\Layerk L}{\Layerk L+\Layerk (N-1)(L+\Memory)-\Layerk_p(L+\Memory)} \nonumber\\
  ~ &=& \frac{1}{N-\theta+(N-1-\theta)\frac{m}{L}},
\end{eqnarray}
which is less than that of the unterminated code. However, the rate loss is negligible for large $L$.

In summary, all code rates of interest in the interval (0,1) can be implemented by adjusting the \emph{repetition degree} $N$ and the \emph{puncturing fraction} $\theta$, all with the encoding structure as shown in Fig.~\ref{BMST_encoder}, where \fbox{P} stands for the optional puncturing.

\subsection{Decoding Algorithm}\label{subsec:Decoding}
Assume that the subcodeword $\boldsymbol{c}^{(t)}$ is modulated using binary phase-shift keying~(BPSK) with 0 and 1 mapped to $+1$ and $-1$, respectively, and transmitted over an AWGN channel, resulting in a received vector $\boldsymbol{y}^{(t)}$ expressed as
\begin{equation}\label{AWGNChannels}
  y_{j}^{(t)} = c_{j}^{(t)} + z_{j}^{(t)},
\end{equation}
for $0 \leq j \leq \Layerk N - \Layerk_p-1$, where $y_{j}^{(t)}$ is the $j$-th component of $\boldsymbol{y}^{(t)}$ and $z_{j}^{(t)}$ is a sample from an independent Gaussian random variable with distribution $\mathcal{N}(0, \sigma^2)$.

\begin{figure}[t]
   \centering
   \includegraphics[angle=270, clip, width=\figwidth]{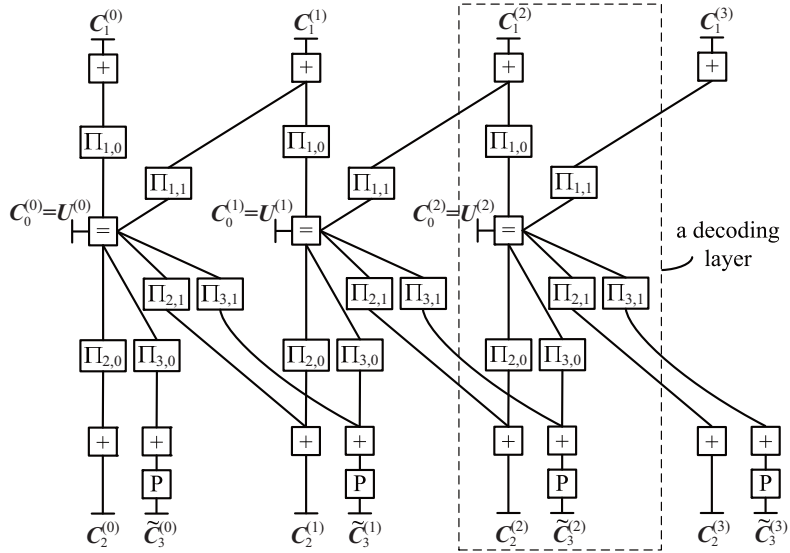}
   \caption{Normal graph of a systematic BMST-R code with $N=4$, $m=1$ and $L=3$.}
   \label{BMST_decoder}
\end{figure}

The decoding algorithm for systematic BMST-R codes can be described as an iterative message processing/passing algorithm over the associated Forney-style factor graph, which is also known as a normal graph~\cite{Forney01}. In the normal graph, edges represent variables and vertices~(nodes) represent constraints. All edges connected to a node must satisfy the specific constraint of the node. A full-edge connects to two nodes, while a half-edge connects to only one node. A half-edge is also connected to a special symbol, called a ``dongle", that denotes coupling to other parts of the transmission system~(say, the channel or the information source)~\cite{Forney01}. Fig.~\ref{BMST_decoder} shows the normal graph of a systematic BMST-R code with $N=4$, $m=1$ and $L=3$. It is indeed a high-level normal graph, where each edge represents a sequence of random variables. There are four types of nodes in the normal graph of the systematic BMST-R code.
\begin{itemize}
  \item \textbf{Node} $\fbox{+}$: All edges~(variables) connected to node $\fbox{+}$ must sum to the all-zero vector. The message updating rule at node $\fbox{+}$ is similar to that of a check node in the factor graph of a binary LDPC code. The only difference is that the messages on the half-edges are obtained from the channel observations.

  \item \textbf{Node} $\fbox{=}$: All edges~(variables) connected to node $\fbox{=}$ must take the same (binary) values. The messages on the half-edges are obtained from both the channel observations and the information source.\footnote{The half-edges~(variables) connected to the information source, which are omitted in Fig.~\ref{BMST_decoder} to avoid confusion and messy plots, are assumed to be independent and uniformly distributed over $\mathbb{F}_2^\Layerk$.} The message updating rule at node $\fbox{=}$ is the same as that of a variable node in the factor graph of a binary LDPC code.

  \item \textbf{Node} \fbox{$\Pi_{i,j}$}: The node \fbox{$\Pi_{i,j}$} represents the $(i,j)$-th interleaver, which interleaves or de-interleaves the input messages.

  \item \textbf{Node} \fbox{P}: Two edges~(variables) connected to node \fbox{P} must satisfy the constraint specified by the puncturing rules.
\end{itemize}

The normal graph of a systematic BMST-R code can be divided into \emph{layers}, where each layer typically consists of a node of type \fbox{=}, $N-1$ nodes of type \fbox{+}, $(m+1)(N-1)$ nodes of type \fbox{$\Pi$}, and a node of type \fbox{P}~(see Fig.~\ref{BMST_decoder}).

Similar to SC-LDPC codes, an iterative sliding window decoding algorithm with decoding delay $d$ performing over a subgraph consisting of $d+1$ consecutive layers can be implemented for systematic BMST-R codes. For each window position, the sliding window decoding algorithm can be implemented using the parallel~(flooding) updating schedule within the decoding window. The first layer in any window is called the {\em target layer}. Decoding proceeds until a fixed number of iterations has been performed or certain given stopping criterion is satisfied, in which case the window shifts to the right by one layer and the symbols corresponding to the target layer shifted out of the window are decoded.

\subsection{Relations of Systematic BMST-R Codes to Existing Codes}\label{subsec:AlgebraicDescription}
From Fig.~\ref{BMST_encoder}, we can see that systematic BMST-R codes resemble the classical RCPC codes~\cite{Hagenauer88}. Evidently, we can start from a rate $1/N$ systematic BMST-R code~(the mother code), where $N$ is as large as required. By puncturing\footnote{If needed, one or more whole branches in Fig.~\ref{BMST_encoder} can be removed.}, one can obtain all code rates of interest from $1/N$ to 1, all of which can be implemented with essentially the same pair of encoder and decoder. The difference between systematic BMST-R codes and RCPC codes is also obvious. The encoding of systematic BMST-R codes is block-oriented and the decoding is typically not implementable by the Viterbi algorithm~\cite{Forney73} due to the huge constraint length induced by the block-oriented encoding process.

Alternatively, systematic BMST-R codes are decodable with a sliding window decoding algorithm, which is similar to SC-LDPC codes. More generally, systematic BMST-R codes can be viewed as a special class of spatially coupled codes, since spatial coupling can be interpreted as introducing memory among successive independent transmissions, where extra edges are allowed to be added during the coupling process~\cite{Huang15JSAC}. In contrast to SC-LDPC codes, which are usually defined by the null space of a sparse parity-check matrix, systematic BMST-R codes are easily described using generator matrices. Further, since the encoder for a systematic BMST-R code is non-recursive, an all-zero tail can be added to drive the encoders to the zero state at the end of the encoding process. This is different from SC-LDPC codes, where the tail is usually non-zero and depends on the encoded information bits~(see Section~IV of~\cite{Pusane08}). As a result, the encoding procedure for systematic BMST-R codes is simpler than for SC-LDPC codes.

When described in terms of generator matrices, systematic BMST codes can also be viewed as a special class of spatially coupled low-density generator-matrix (SC-LDGM) codes~\cite{Aref12,Kumar_IT14}. However, as an ensemble, systematic BMST-R codes are different from SC-LDGM codes. SC-LDGM code ensembles are usually defined in terms of their node distributions, while systematic BMST-R code ensembles are defined in terms of their interleavers~(see Fig.~\ref{BMST_encoder}).

As another evidence that systematic BMST-R codes are different from existing codes, we would like to emphasize that systematic BMST-R codes have a simple lower bound on the BER performance, as described in the next section.

\section{Performance and Complexity Analysis}\label{SecIII}
A reasonable criterion for a construction to be good is its ability to make trade-offs between complexity and performance. Specifically, if the error performance required by the user is relaxed or, if the gap between the code rate and the capacity is more tolerant, the encoding/decoding complexity should be reduced. In this section, we will find a relation of the performance to the complexity for {\em terminated} systematic BMST-R codes. We start with a general systematic linear block code.

\subsection{Basic Notations of Systematic Linear Block Codes}\label{subsec:Notation}
A binary linear block code $\mathcal{C}[\Alln,\Allk]$ is a $\Allk$-dimensional subspace of $\mathbb{F}_2^\Alln$. An encoding algorithm can be described simply by
\begin{equation}\label{encoding}
\begin{array}{cc}
  \phi:~\mathbb{F}_2^\Allk \rightarrow \mathbb{F}_2^\Alln\\
  ~~~~~~~~~{\boldsymbol u} \rightarrow {\boldsymbol c}={\boldsymbol u}{\boldsymbol G},
  \end{array}
\end{equation}
where ${\boldsymbol u} \in \mathbb{F}_2^\Allk$ is the information vector, ${\boldsymbol c}$ is the associated codeword, and ${\boldsymbol G}$ is a generator matrix of size $\Allk \times \Alln$ with rank of $\Allk$. Define
\begin{equation}\label{C_1i}
\mathcal{C}_{1,i} \stackrel{\Delta}{=} \{{\boldsymbol{c} = \boldsymbol{u}\boldsymbol{G}: ~u_i = 1}\}.
\end{equation}
Let $d_{\min,i}$ be the minimum Hamming weight of $\mathcal{C}_{1,i}$, i.e.,
\begin{equation}\label{d_mini}
  d_{\min,i}  \stackrel{\Delta}{=}  \min\limits_{\boldsymbol{c} \in \mathcal{C}_{1,i}} W_H(\boldsymbol{c}),
\end{equation}
where $W_H(\cdot)$ represents the Hamming weight. Obviously, the minimum Hamming weight $d_{\min}$ of the linear block code $\mathcal{C}$ can be given by
\begin{equation}\label{dmin}
  d_{\min}= \min\limits_i d_{\min,i}.
\end{equation}

Assume that the codeword $\boldsymbol{c}$ is modulated using BPSK and transmitted over an AWGN channel, resulting in a received vector $\boldsymbol{y}$. A decoding algorithm is defined as a mapping
\begin{equation}\label{decoding}
  \begin{array}{cc}
  \psi:~\mathcal{Y}^\Alln \rightarrow \mathbb{F}_2^\Allk \\
  ~~~~~~~~~~~{\boldsymbol y} \rightarrow \hat{\boldsymbol u}=\psi({\boldsymbol y}),
  \end{array}
\end{equation}
where $\mathcal{Y} \subset \mathbb{R}$. Given the signal mapping $0\rightarrow +1$ and $1\rightarrow -1$, the SNR is given by $10 \log_{10} (1/\sigma^2)$ in dB, where $\sigma^2$ is the variance of the noise.

Suppose that $\boldsymbol{U}$ is distributed uniformly at random over $\mathbb{F}_2^\Allk$. Let $E \stackrel{\Delta}{=} \{\hat{\boldsymbol{U}} \neq {\boldsymbol{U}}\}$ be the error event that the decoder output $\hat{\boldsymbol{U}}$ is not equal to the encoder input vector $\boldsymbol{U}$, and let $E_{i} \stackrel{\Delta}{=} \{ {\boldsymbol{\hat U}}_i \neq \boldsymbol{U}_i\}$ be the error event that the $i$-th estimated bit $\hat{\boldsymbol{U}}_i$ at the decoder is not equal to the $i$-th input bit $\boldsymbol{U}_i$. Obviously, $E = \bigcup\limits_{0\leq i \leq \Allk-1} E_{i}$. Then, under the given decoding algorithm $\psi$, we can define frame error probability
\begin{equation}\label{PeDef0}
    {\rm FER}_{\psi} \stackrel{\Delta}{=} {\rm Pr}\{E\},
\end{equation}
and bit-error probability
\begin{equation}\label{PbDef0}
    {\rm BER}_{\psi} \stackrel{\Delta}{=} \frac{1}{\Allk} \sum_{0\leq i\leq \Allk-1} {\rm Pr}\{ E_{i}\}.
\end{equation}
From the definitions of BER and FER, we have
    \begin{equation}\label{Proof1}
        {\rm FER_{\psi}} = {\rm Pr}\left\{ \bigcup_i E_{i}\right\} \geq \max_i {\rm Pr}\{ E_{i}\} \geq {\rm BER_{\psi}}.
    \end{equation}
We also have
    \begin{equation}\label{Proof2}
        {\rm FER_{\psi}} = {\rm Pr}\left\{ \bigcup_i E_{i}\right\} \leq
        \sum_{0\leq i\leq \Allk-1} {\rm Pr}\{ E_{i}\} = \Allk~{\rm BER}_{\psi}.
    \end{equation}
Thus, we have
    \begin{equation}\label{Eq1}
        {\rm BER_{\psi}} \leq {\rm FER_{\psi}} \leq \Allk~{\rm BER_{\psi}}.
    \end{equation}

The maximum-likelihood~(ML) decoding algorithm selects a codeword $\hat{\boldsymbol{c}}$ such that $f(\boldsymbol{y}|\hat{\boldsymbol{c}}) \geq f({\boldsymbol{y}}|{\boldsymbol{c}})$ for all codewords ${\boldsymbol{c}}$. If ties happen, the ML decoding algorithm can randomly select one candidate as the decoder output. Since $\boldsymbol{U}$ is distributed uniformly at random over $\mathbb{F}_2^{\Allk}$, the ML decoding algorithm is optimal in the sense that it minimizes the FER. To minimize the BER, the MAP decoding algorithm computes
\begin{eqnarray}\label{bit}
    {\rm Pr}(U_i=0|{\boldsymbol{y}}) &=& \frac{\sum\limits_{\boldsymbol{u}:u_i=0}{\rm Pr}(\boldsymbol{u})f({\boldsymbol{y}}|\boldsymbol{u}\boldsymbol{G})}{\sum\limits_{\boldsymbol{u}\in \mathbb{F}_2^{\Allk}}{\rm Pr}(\boldsymbol{u})f({\boldsymbol{y}}|\boldsymbol{u}\boldsymbol{G})},
\end{eqnarray}
for all $i$. For each $i = 0, 1,\cdots,{\Allk}-1$, the MAP decoding algorithm outputs ${\hat u_i}=0$ if ${\rm Pr}(U_i=0|{\boldsymbol{y}}) > 0.5$ and ${\hat u_i}=1$ otherwise.

The IRWEF of a systematic block code can be given as~\cite{Benedetto96}
\begin{equation}\label{eq:IRWEF}
    A \left( X, Y \right) \triangleq \sum_{i,j} A_{i,j} X^{i} Y^{j},
\end{equation}
where $X$, $Y$ are two dummy variables and $A_{i,j}$ denotes the number of codewords having input~(information bits) weight $i$ and redundancy~(parity check bits) weight $j$. The IRWEF can also be written in a more compact form as
\begin{equation}\label{eq:IRWEFCompact}
    A \left( X, Y \right) = \sum_{i} A_{i}\left( Y \right) X^{i},
\end{equation}
where
\begin{equation}\label{eq:CRWEF}
    A_i \left( Y \right) \triangleq \sum_{j} A_{i,j} Y^{j}
\end{equation}
is the conditional redundancy weight enumerating function~(CRWEF), which enumerates redundancy weight for a given input weight $i$.


\subsection{Upper Bound on BER Performance}\label{subsec:UpperBound}
Since MAP decoding is optimal in the sense that it minimizes the BER, an upper bound on BER performance under any decoding algorithm is applicable to the MAP decoding algorithm. In the following, we consider a suboptimal list decoding algorithm.

\vspace{0.15cm}
\begin{algorithm}{A List Decoding Algorithm for the Purpose of Performance Analysis}\label{alg:ListDecoding}
\begin{enumerate}
  \item Make hard decisions on the information part of the received vector $\boldsymbol{y}$, resulting in a vector $\hat{\boldsymbol{y}}$ of length $\Allk$. Then the channel becomes a memoryless binary symmetric channel~(BSC) with cross probability
        \begin{equation}\label{epsilon}
          \varepsilon \stackrel{\Delta}{=}Q\left(\frac{1}{\sigma}\right).
        \end{equation}

  \item List all sequences of length $\Allk$ within the Hamming sphere with center at $\hat{\boldsymbol{y}}$ of radius $r^*\geq 0$. The resulting list is denoted as $\mathcal{L}_{\boldsymbol{y}}$.

  \item Encode each sequence in $\mathcal{L}_{\boldsymbol{y}}$ by the encoding algorithm of the systematic code, resulting in a list of codewords, denoted as $\mathcal{L}_{\boldsymbol{c}}$.

  \item Find the codeword ${\boldsymbol{c}}^*\in\mathcal{L}_{\boldsymbol{c}}$ that is closest to $\boldsymbol{y}$. Output the information part $\hat{\boldsymbol{u}}$ of ${\boldsymbol{c}^*}$ as the decoding result.
\end{enumerate}
\end{algorithm}
\vspace{0.15cm}

The above list decoding algorithm is similar to but different from the algorithm presented in~\cite{Ma13Bound}. The {\em list region} in~\cite{Ma13Bound} is an $n$-dimensional Hamming sphere with center at the hard decision of the whole received sequence, while the list region here is a $k$-dimensional Hamming sphere with center at the hard decision of the information part of the received sequence. By analyzing the BER performance of the proposed list decoding algorithm, we have the following theorem.

%

\vspace{0.15cm}
\begin{theorem}\label{TheoremUpperBound}
For any integer $r^* \geq 0$, the bit-error probability of systematic codes under MAP decoding is upper-bounded by
\ifCLASSOPTIONonecolumn
\begin{equation}\label{UpperBound}
    {\rm BER_{MAP}}\leq \sum\limits_{i \leq 2r^*}\frac{i}{\Allk}\left( \sum\limits_{j}A_{i,j}\Qfun{\frac{\sqrt{i+j}}{\sigma}}\right)
    + \sum_{i=r^*+1}^{\Allk}\frac{\min\{i+r^*,\Allk\}}{\Allk}\binom{\Allk}{i} \varepsilon^i (1-\varepsilon)^{\Allk-i}.
\end{equation}
\fi
\ifCLASSOPTIONtwocolumn
\begin{align}\label{UpperBound}
    {\rm BER_{MAP}}\leq \sum\limits_{i \leq 2r^*}\frac{i}{\Allk}\left( \sum\limits_{j}A_{i,j}\Qfun{\frac{\sqrt{i+j}}{\sigma}}\right)~~~~~~\nonumber\\
    + \sum_{i=r^*+1}^{\Allk}\frac{\min\{i+r^*,\Allk\}}{\Allk}\binom{\Allk}{i} \varepsilon^i (1-\varepsilon)^{\Allk-i}.
\end{align}
\fi
\end{theorem}

\begin{IEEEproof}
Consider the list decoding algorithm~(Algorithm~\ref{alg:ListDecoding}). The decoding error occurs in two cases under the assumption that the all-zero codeword is transmitted.
\begin{enumerate}
  \item The all-zero sequence of length $\Allk$ is not in the list $\mathcal{L}_{\boldsymbol{y}}$, i.e., the hard-decisions have $i \geq r^*+1$ errors. In this case, the decoder output has at most $i+r^*$ erroneous bits. Hence, the bit-error probability, denoted as $p_1$, is upper-bounded by
        \begin{equation}\label{p1}
            p_1 \leq \sum_{i=r^*+1}^{\Allk}\frac{\min\{i+r^*,\Allk\}}{\Allk}\binom{\Allk}{i} \varepsilon^i (1-\varepsilon)^{\Allk-i}.
        \end{equation}

  \item The all-zero sequence of length $\Allk$ is in the list $\mathcal{L}_{\boldsymbol{y}}$, but the all-zero codeword $\boldsymbol{c}^{(0)}$ is not the closest one to $\boldsymbol{y}$. In this case, the bit-error probability, denoted as $p_2$, is upper-bounded by
        \begin{equation}\label{p2}
            p_2 \leq \sum\limits_{i \leq 2r^*}\frac{i}{\Allk}\left( \sum\limits_{j}A_{i,j}\Qfun{\frac{\sqrt{i+j}}{\sigma}}\right).
        \end{equation}
\end{enumerate}

In summary, for any given radius $r^*$, we have
\ifCLASSOPTIONonecolumn
\begin{equation}\label{List}
    {\rm BER_{List}}\leq \sum\limits_{i \leq 2r^*}\frac{i}{\Allk}\left( \sum\limits_{j}A_{i,j}\Qfun{\frac{\sqrt{i+j}}{\sigma}}\right)
    + \sum_{i=r^*+1}^{\Allk}\frac{\min\{i+r^*,\Allk\}}{\Allk}\binom{\Allk}{i} \varepsilon^i (1-\varepsilon)^{\Allk-i}.
\end{equation}
\fi
\ifCLASSOPTIONtwocolumn
\begin{align}\label{List}
    {\rm BER_{List}}\leq \sum\limits_{i \leq 2r^*}\frac{i}{\Allk}\left( \sum\limits_{j}A_{i,j}\Qfun{\frac{\sqrt{i+j}}{\sigma}}\right)~~~~~~\nonumber\\
    + \sum_{i=r^*+1}^{\Allk}\frac{\min\{i+r^*,\Allk\}}{\Allk}\binom{\Allk}{i} \varepsilon^i (1-\varepsilon)^{\Allk-i}.
\end{align}
\fi
Combining~(\ref{List}) and the fact that ${\rm BER_{MAP}} \leq {\rm BER_{List}}$, we complete the proof.
\end{IEEEproof}

From Theorem~\ref{TheoremUpperBound}, we have the following three corollaries.

\vspace{0.15cm}
\begin{corollary}\label{Corollary1}
\begin{align}
    {\rm BER_{MAP}} \leq \sum\limits_{i=1}^{\Allk}\frac{i}{\Allk}\left( \sum\limits_{j}A_{i,j}\Qfun{\frac{\sqrt{i+j}}{\sigma}}\right).
\end{align}
\end{corollary}

\begin{IEEEproof}
It can be proved by simply setting $r^*=\Allk$ in~(\ref{UpperBound}).
\end{IEEEproof}

\vspace{0.15cm}
\begin{corollary}\label{Corollary2}
\begin{align}\label{CorollaryMAP2}
    {\rm BER_{MAP}} \leq Q\left(\frac{1}{\sigma}\right).
\end{align}
\end{corollary}

\begin{IEEEproof}
By simply setting $r^*=0$ in~(\ref{UpperBound}), we have
\begin{eqnarray}\label{Corollary2Proof}
    {\rm BER_{MAP}} &\leq& \sum_{i=1}^{\Allk}\frac{i}{\Allk}\binom{\Allk}{i} \varepsilon^i (1-\varepsilon)^{\Allk-i} \nonumber \\
    &=& \varepsilon = Q\left(\frac{1}{\sigma}\right).
\end{eqnarray}
\end{IEEEproof}

\vspace{0.15cm}
\begin{corollary}\label{Corollary3}
Assuming that we know only the truncated IRWEF $\{A_{i,j}$, $0 \leq i \leq T\}$ of systematic codes, we have
\ifCLASSOPTIONonecolumn
\begin{equation}\label{MAP_UpperBound}
    {\rm BER_{MAP}} \leq \min\limits_{0\leq r^*\leq T/2}\Bigg\{  \sum\limits_{i \leq 2r^*}\frac{i}{\Allk}\left( \sum\limits_{j}A_{i,j}\Qfun{\frac{\sqrt{i+j}}{\sigma}}\right)
    + \sum_{i=r^*+1}^{\Allk}\frac{\min\{i+r^*,\Allk\}}{\Allk}\binom{\Allk}{i} \varepsilon^i (1-\varepsilon)^{\Allk-i} \Bigg\}.
\end{equation}
\fi
\ifCLASSOPTIONtwocolumn
\begin{align}\label{MAP_UpperBound}
    {\rm BER_{MAP}} \leq \min\limits_{0\leq r^*\leq T/2}\Bigg\{  \sum\limits_{i \leq 2r^*}\frac{i}{\Allk}\left( \sum\limits_{j}A_{i,j}\Qfun{\frac{\sqrt{i+j}}{\sigma}}\right)\nonumber\\
    + \sum_{i=r^*+1}^{\Allk}\frac{\min\{i+r^*,\Allk\}}{\Allk}\binom{\Allk}{i} \varepsilon^i (1-\varepsilon)^{\Allk-i} \Bigg\}.
\end{align}
\fi
\end{corollary}

\begin{IEEEproof}
It is obvious and omitted here.
\end{IEEEproof}

\textbf{Remarks.}~Corollary~\ref{Corollary1} is the well-known union bound, while Corollary~\ref{Corollary2} is almost trivial, which can be easily understood by noting that setting $r^*=0$ in  Algorithm~\ref{alg:ListDecoding} is equivalent to taking directly the hard decisions $\hat{\boldsymbol{y}}$ as the decoding result $\hat{\boldsymbol{u}}$~(one of the simplest sub-optimal decoding algorithms). Given that only the truncated IRWEF is available, Corollary~\ref{Corollary3} is the tightest upper bound of this type.

\subsection{Lower Bound on BER Performance}\label{subsec:LowerBound}

There exist several lower bounds on FER under ML decoding~\cite{Seguin98,Cohen04,Sason06,Behnamfar07}. However, lower bounds on BER are rarely mentioned in the literature. Any lower bound on ${\rm FER_{ML}}$ can be adapted to a lower bound on BER by noticing that ${\rm BER_{ML}} \geq \frac{1}{\Allk}~{\rm FER_{ML}}$ from~(\ref{Eq1}). The simplest lower bound on FER under ML decoding over AWGN channels is given by
\begin{equation}\label{ML_propoFER}
    {\rm FER_{ML}} \geq ~Q\left(\frac{\sqrt{d_{\min}}}{\sigma}\right),
\end{equation}
which leads to
\begin{equation}\label{ML_propo}
    {\rm BER_{ML}} \geq \frac{1}{\Allk}~{\rm FER_{ML}} \geq \frac{1}{{\Allk}}~Q\left(\frac{\sqrt{d_{\min}}}{\sigma}\right).
\end{equation}
Logically, it is not safe to conclude from the above derivation that the lower bound~(\ref{ML_propo}) applies to MAP decoding. This is subtle due to the fact that ML decoding is not optimal for minimizing the bit-error probability. In the following, we will show that the lower bound on ${\rm BER_{ML}}$~(\ref{ML_propo}) is indeed a lower but usually loose bound on ${\rm BER_{MAP}}$ by proving an improved lower bound.\footnote{A slightly surprising fact is that no lower bound on ${\rm BER_{MAP}}$ was found with proof in the literature.} To see the looseness of the lower bound, we consider the following toy example.

Let $\mathcal{A} = \{00, 10\}$ with $d_{\rm min} = 1$ and $\mathcal{B} = \{00, 11\}$ with $d_{\rm min} = 2$ be two codes. Define $\mathcal{C} = \mathcal{A} \times \mathcal{B}^{9999}$, whose codewords are in a Cartesian product form $(\boldsymbol{c}_0, \boldsymbol{c}_1, \cdots, \boldsymbol{c}_{9999})$, where $\boldsymbol{c}_0 \in \mathcal{A}$ and $\boldsymbol{c}_i \in \mathcal{B}$ for $1 \leq i \leq 9999$. Obviously, the code $\mathcal{C}$ has minimum Hamming weight $d_{\rm min} = 1$. However, for BPSK modulation over an AWGN channel, the BER for the code $\mathcal{C}$ is dominated by the code $\mathcal{B}$ rather than the code $\mathcal{A}$. To be precise,
\begin{eqnarray}\label{MAP_ML}
  {\rm BER_{MAP}} = \frac{1}{10000}Q\left(\frac{1}{\sigma}\right) + \frac{9999}{10000}Q\left(\frac{\sqrt{2}}{\sigma}\right),
\end{eqnarray}
which implies that the lower bound ${\rm BER_{MAP}} \geq \frac{1}{10000}Q\left(\frac{1}{\sigma}\right)$ can be very loose in the low SNR region. In the following we present an improved lower bound under MAP decoding.

\vspace{0.15cm}
\begin{theorem}\label{MAPLowerBound}
The bit-error probability for the linear block code $\mathcal{C}$ under MAP decoding can be lower-bounded by
\begin{equation}\label{BitBound}
    {\rm BER_{MAP}} \geq \frac{1}{\Allk} \sum\limits_{i=0}^{\Allk-1} Q \left(\frac{\sqrt{d_{\min,i}}}{\sigma}\right).
\end{equation}
\end{theorem}

\begin{IEEEproof}
It suffices to prove that ${\rm Pr}\{E_i\} \geq Q \left(\frac{\sqrt{d_{\min,i}}}{\sigma}\right)$ for each given $i$~$(0\leq i \leq \Allk-1)$. Let $\boldsymbol{c}^{(1)} \in \mathcal{C}_{1,i}$ be a codeword such that $ d_{\min,i}= W_H(\boldsymbol{c}^{(1)})$. There must exist an invertible matrix $\boldsymbol{T}$ of size $\Allk\times \Allk$ such that $\boldsymbol{G}=\boldsymbol{T}\widetilde{\boldsymbol{G}}$ with the first row of $\widetilde{\boldsymbol{G}}$ being $\boldsymbol{c}^{(1)}$. Assume $\boldsymbol{U} \in \mathbb{F}_2^{\Allk}$ be the information vector and $\boldsymbol{C}=\boldsymbol{U}\boldsymbol{G}$ be the codeword to be transmitted. Define $\boldsymbol{V}=\boldsymbol{U}\boldsymbol{T}$. The MAP decoder for a binary linear block code computes ${\rm Pr}\left\{ u_i|\boldsymbol{y} \right\}$. We know that if ${\rm Pr}\{u_i | \boldsymbol{y}\} > 0.5$, the decoding output is correct for this considered bit. In the meanwhile, we assume a \emph{genie-aided decoder}, which computes ${\rm Pr}\{u_i | \boldsymbol{y}, \boldsymbol{v}'\}$ with $\boldsymbol{v}' = (v_1, v_{2}, \cdots, v_{\Allk-1})$ available. Likewise, if ${\rm Pr}\{u_i | \boldsymbol{y}, \boldsymbol{v}'\} > 0.5$, the decoding output is correct for this considered bit. For a specific $\boldsymbol{u}$ and $\boldsymbol{y}$, it is possible that ${\rm Pr}\{u_i | \boldsymbol{v}', \boldsymbol{y}\} < {\rm Pr}\{u_i | \boldsymbol{y}\}$. However, the expectation
\begin{eqnarray}\label{eq:MutualINfo}
\mathbb{E}\left[ \log \frac{{\rm Pr}\{u_i | \boldsymbol{v}',
\boldsymbol{y}\}}{{\rm Pr}\{u_i | \boldsymbol{y}\}} \right] = I
\left(U_i; \boldsymbol{V}' | \boldsymbol{Y}\right)
\geq 0,
\end{eqnarray}
where $I \left(U_i; \boldsymbol{V}'  | \boldsymbol{Y}\right)$ is the conditional mutual information. This implies that the genie-aided decoder performs statistically no worse than the MAP decoder of the binary linear block code. Under the condition that $\boldsymbol{v}'$ is available, there exist only two codewords whose Hamming distance is $d_{\min,i}$. Thus, the bit-error probability with the genie-aided decoder for the binary-input AWGN channels is ${\rm Pr}\{E_i\}_{\rm Genie}=Q \left(\frac{\sqrt{d_{\min,i}}}{\sigma}\right)$. It follows that
\begin{equation}\label{Theorem2_eq}
    {\rm Pr}\{E_i\} \geq {\rm Pr}\{E_i\}_{\rm Genie}= Q \left(\frac{\sqrt{d_{\min,i}}}{\sigma}\right).
\end{equation}
\end{IEEEproof}

\textbf{Remarks.}~Theorem~\ref{MAPLowerBound} also applies to {\em non-systematic} linear block codes. However, it does not apply to non-linear codes, indicating that the proof is not that simple as considering only the two closest codewords.

From Theorem~\ref{MAPLowerBound}, we have the following three corollaries.

\begin{corollary}\label{ML_LowerBound2MAP}
\begin{eqnarray}\label{ML_LowerBound2MAP_eq}
    {\rm BER_{MAP}} \geq \frac{1}{{\Allk}}~Q\left(\frac{\sqrt{d_{\min}}}{\sigma}\right).
\end{eqnarray}
\end{corollary}

\begin{IEEEproof}
Combining~(\ref{dmin}) and Theorem~\ref{MAPLowerBound}, we have
\begin{equation}
    {\rm BER_{MAP}} \geq \frac{1}{\Allk} \sum\limits_{i=0}^{\Allk-1} Q \left(\frac{\sqrt{d_{\min,i}}}{\sigma}\right)  \geq \frac{1}{{\Allk}}~Q\left(\frac{\sqrt{d_{\min}}}{\sigma}\right).
\end{equation}
\end{IEEEproof}

\begin{corollary}\label{Cyclic_LowerBound}
If a code\footnote{A cyclic code can be such an example.} has the property that $d_{\min,i}=d_{\min}$ for all $i$, we have
\begin{eqnarray}\label{Pb_AWGN}
    {\rm BER_{MAP}} \geq Q\left(\frac{\sqrt{d_{\min}}}{\sigma}\right).
\end{eqnarray}
\end{corollary}

\begin{IEEEproof}
It is obvious and omitted here.
\end{IEEEproof}

\begin{corollary}\label{LDGM_LowerBound}
If the row weights of the generator matrix $\boldsymbol{G}$ for a linear block code $\mathcal{C}$ are $w_0, w_1, \cdots,w_{{\Allk}-1}$, we have
\begin{eqnarray}\label{LDGM_LowerBound_sq}
    {\rm BER_{MAP}} \geq \frac{1}{\Allk} \sum\limits_{i=0}^{\Allk-1} Q \left(\frac{\sqrt{w_{i}}}{\sigma}\right).
\end{eqnarray}
\end{corollary}

\begin{IEEEproof}
This can be proved by noting that $d_{\min,i} \leq w_{i}$ and that $Q(x)$ is a decreasing function.
\end{IEEEproof}

\textbf{Remarks.}~Corollary~\ref{ML_LowerBound2MAP} shows that the lower bound~(\ref{ML_propo}) on ${\rm BER_{ML}}$ is also a lower bound on the ${\rm BER_{MAP}}$, while Corollary~\ref{Cyclic_LowerBound} indicates that the lower bound~(\ref{ML_propo}) can be very loose. Corollary~\ref{LDGM_LowerBound} indicates that an LDGM code may have a higher error floor compared to an LDPC code, since the generator matrix for an LDPC code is typically high-density.

\subsection{Applications to Systematic BMST-R Codes}\label{subsec:ApplicationBound}
To apply the derived bounds to systematic BMST-R codes, we need calculate the IRWEF. For systematic BMST-R codes, we have
\ifCLASSOPTIONonecolumn
\begin{eqnarray}
 A(X, Y)&=& \sum_{i,j} A_{i,j} X^{i} Y^{j} \nonumber \\
        &=& \sum_{\boldsymbol{u}}X^{W_H(\boldsymbol{u})} \prod\limits_{t = 0}^{L+m-1} \left( Y^{W_H(\widetilde{\boldsymbol{c}}_{N-1}^{(t)})} \prod\limits_{i=1}^{N-2} Y^{W_H(\boldsymbol{c}_i^{(t)})} \right) \nonumber \\
        &=& \sum_{\boldsymbol{u}}\prod_{t = 0}^{L+m-1} \left(X^{W_H(\boldsymbol{u}^{(t)})}Y^{W_H(\widetilde{\boldsymbol{c}}_{N-1}^{(t)})}\prod\limits_{i=1}^{N-2} Y^{W_H(\boldsymbol{c}_i^{(t)})}\right),
\end{eqnarray}
\fi
\ifCLASSOPTIONtwocolumn
\begin{eqnarray}
A(X, Y)=\sum_{i,j} A_{i,j} X^{i} Y^{j} ~~~~~~~~~~~~~~~~~~~~~~~~~~~~~~~~~~~~\nonumber \\
        = \sum_{\boldsymbol{u}}X^{W_H(\boldsymbol{u})} \prod\limits_{t = 0}^{L+m-1} \left( Y^{W_H(\widetilde{\boldsymbol{c}}_{N-1}^{(t)})} \prod\limits_{i=1}^{N-2} Y^{W_H(\boldsymbol{c}_i^{(t)})} \right) ~~\nonumber \\
        = \sum_{\boldsymbol{u}}\prod_{t = 0}^{L+m-1} \left(X^{W_H(\boldsymbol{u}^{(t)})}Y^{W_H(\widetilde{\boldsymbol{c}}_{N-1}^{(t)})}\prod\limits_{i=1}^{N-2} Y^{W_H(\boldsymbol{c}_i^{(t)})}\right),
\end{eqnarray}
\fi
where the summation is over all possible data sequences $\boldsymbol{u}$ with $\boldsymbol{u}^{(t)} = \boldsymbol{0}$ for $t \geq L$. Since it is a sum of products, $A(X, Y)$ can be computed in principle by a trellis-based algorithm over the polynomial ring. For specific interleavers, the trellis has a state space of size $2^{m\Layerk}$, which makes the computation intractable for large $m\Layerk$. To circumvent this issue, we turn to an ensemble of systematic BMST-R codes by assuming that all the interleavers~(see Fig.~\ref{BMST_encoder} for reference) are chosen at each time independently and uniformly at random, and that $\widetilde{\boldsymbol{c}}_{N-1}^{(t)}$ is obtained by randomly puncturing $\Layerk_p$ of $\Layerk$ bits in $\boldsymbol{c}_{N-1}^{(t)}$. With the assumption that all interleavers are uniform interleavers~\cite{Benedetto96}, we can see that $W_H(\boldsymbol{c}_{i}^{(t)})$, for $1 \leq i \leq N-1$, is a random variable which depends {\em only} on the Hamming weights $\{W_H(\boldsymbol{u}^{(t-j)}), 0 \leq j \leq m\}$. This admits a reduced-complexity trellis representation of the average IRWEF of the defined systematic BMST-R code ensemble.


The trellis is time-invariant. At stage $t$, the trellis has $(\Layerk +1)^{m}$ states, each of which corresponds to a vector of Hamming weights $\boldsymbol{p} = \left(W_H(\boldsymbol{u}^{(t-1)}),W_H(\boldsymbol{u}^{(t-2)}),\cdots,W_H(\boldsymbol{u}^{(t-m)})\right)$. A state $\boldsymbol{p}$ at stage $t$ and a state $\boldsymbol{q}$ at stage $t+1$ are connected~(with a branch denoted by $\boldsymbol{p}\rightarrow \boldsymbol{q}$) if and only if $p_{j} = q_{j+1}$ for $0 \leq j \leq m-2$, where $p_j$ and $q_j$ are the $j$-th components of $\boldsymbol{p}$ and $\boldsymbol{q}$, respectively. Evidently, emitting from~(or entering into) each state, there are $\Layerk +1$ branches. Associated with a branch $\boldsymbol{p}\rightarrow \boldsymbol{q}$ are a deterministic input weight $q_0$ but a random redundancy weight due to the existence of random interleavers. The weight distribution of the parity check vector $\boldsymbol{c}_{1}^{(t)}$ is given by
\begin{equation}
\gamma_{\boldsymbol{p}\rightarrow \boldsymbol{q}}=\sum\limits_{r=0}^{\Layerk} f(r|\boldsymbol{p},q_0) Y^{r},
\end{equation}
where $f(r|\boldsymbol{p},q_0)$ is interpreted as the probability of current outputs $\boldsymbol{c}_{1}^{(t)}$ having weight $r$ given that the weight vector of previous $m$ input blocks $\left(W_H(\boldsymbol{u}^{(t-1)}),W_H(\boldsymbol{u}^{(t-2)}),\cdots,W_H(\boldsymbol{u}^{(t-m)})\right)=\boldsymbol{p}$ and the current input weight $W_H(\boldsymbol{u}^{(t)}) = q_0$. By symmetry, it is easy to see that the weight distribution of $\boldsymbol{c}_{i}^{(t)}$ for $1 \leq i \leq N-2$ is the same as $\gamma_{\boldsymbol{p}\rightarrow \boldsymbol{q}}$. Since the parity check vector $\widetilde{\boldsymbol{c}}_{N-1}^{(t)}$ is obtained by randomly puncturing $\Layerk_p$ of $\Layerk$ bits in $\boldsymbol{c}_{N-1}^{(t)}$, the weight distribution of $\widetilde{\boldsymbol{c}}_{N-1}^{(t)}$ is given by\footnote{By a general definition, the binomial coefficient $\binom{n}{k}$ is equal to zero for $k<0$ or $k>n$.}
\begin{equation}
\widetilde{\gamma}_{\boldsymbol{p}\rightarrow \boldsymbol{q}}=\sum\limits_{r=0}^{\Layerk} \left( f(r|\boldsymbol{p},q_0) \sum\limits_{w=0}^{\Layerk} \frac{\binom{r}{w}\binom{\Layerk-r}{\Layerk_p-w}}{\binom{\Layerk}{\Layerk_p}}Y^{r-w} \right).
\end{equation}

To calculate the probability $f(r|\boldsymbol{p},q_0)$, we define $g\left( r|p,q \right)$ as the probability that a vector of length $\Layerk$ has weight $r$, given that the vector is obtained by superimposing two randomly interleaved vectors of~(respective) weights $p$ and $q$. Define $w \stackrel{\Delta}{=} p+q-r$. Following the same lines as the method in Section~IV-B of~\cite{Ma15}, the probability $g\left( r|p,q \right)$ is given by
\begin{equation}\label{Prob_pqr}
  g\left( r|p,q \right) =
          \left\{ \begin{array}{cc}
            \frac{ \binom{q}{w/2}\binom{\Layerk-q}{p-w/2} }{ \binom{\Layerk}{p} }, &{\rm if}~w~{\rm is~even},\\
            0, &{\rm otherwise}.
        \end{array}\right.
\end{equation}
Then, $f(r|\boldsymbol{p},q_0)$ can be calculated as described in Algorithm~\ref{alg:f_pqr}.

\newpage
\begin{algorithm}{Computing the probability $f(r|\boldsymbol{p},q_0)$}\label{alg:f_pqr}
\begin{enumerate}
  \item Initialize a vector $\boldsymbol{\alpha}^{(0)}$ of length $\Layerk+1$ such that its components are all zero except that the $q_0$-th component is 1.
  \item For $j = 0$, $1$, $\cdots$, $m-1$, compute
        \begin{equation}
          \alpha_{i}^{(j+1)}=\sum\limits_{\ell=0}^{\Layerk} \alpha_{\ell}^{(j)} g\left( i|\ell,p_{j} \right),
          \nonumber
        \end{equation}
        for $0 \leq i \leq \Layerk$, where $\alpha_{\ell}^{(j)}$ is the $\ell$-th component of $\boldsymbol{\alpha}^{(j)}$ and $p_{j}$ is the $j$-th component of $\boldsymbol{p}$.
  \item We have $f(r|\boldsymbol{p},q_0)=\alpha_{r}^{(m)}$ for $r = 0$, $1$, $\cdots$, $\Layerk$.
\end{enumerate}
\end{algorithm}
\vspace{0.15cm}

Finally, $A(X,Y)$ can be calculated recursively by performing a forward trellis-based algorithm~\cite{Ma03} over the polynomial ring in Algorithm~\ref{alg:IOWEF}.

\vspace{0.15cm}
\begin{algorithm}{Computing IRWEF of Systematic BMST-R Codes}\label{alg:IOWEF}
\begin{enumerate}
  \item Initialize $\beta_0(\boldsymbol{p})=1$ if $\boldsymbol{p}$ is the all-zero state; otherwise, initialize $\beta_0(\boldsymbol{p})=0$.
  \item For $t = 0$, $1$, $\cdots$, $L+m-1$, for each state $\boldsymbol{q}$,
        \begin{equation}
          \beta_{t+1}(\boldsymbol{q})=\sum_{\boldsymbol{p}:\boldsymbol{p}\rightarrow \boldsymbol{q}} \binom{\Layerk}{q_0}X^{q_0} \widetilde{\gamma}_{\boldsymbol{p}\rightarrow \boldsymbol{q}} \left(\gamma_{\boldsymbol{p}\rightarrow \boldsymbol{q}}\right)^{N-2} \beta_t(\boldsymbol{p}),
          \nonumber
        \end{equation}
        where $q_0 \in \{0,1,\cdots,\Layerk\}$ is the first component of $\boldsymbol{q}$.
  \item At time $L+m$, we have $A(X,Y)=\beta_{L+m}(\boldsymbol{0})$.
\end{enumerate}
\end{algorithm}\vspace{0.15cm}

\textbf{Remarks.}~The summation for Step~2) in Algorithm~\ref{alg:IOWEF} is over $\Layerk+1$ possible states $\boldsymbol{p}$ for a given state $\boldsymbol{q}$. The computation of Algorithm~\ref{alg:IOWEF} becomes more complicated and even intractable for large $m$ and/or $\Layerk$ due to the huge number of trellis states $(\Layerk +1)^{m}$. Fortunately, as shown in Section~\ref{subsec:UpperBound}, we can calculate bounds by the use of a truncated IRWEF, which can be obtained by removing certain states and branches from the trellis. Specifically, for a given truncating parameter $T$ which corresponds to the maximum input weight, we remove all the branches $\boldsymbol{p}\rightarrow \boldsymbol{q}$ with $q_0 + \sum\limits_{j=0}^{m-1} p_j > T$ and keep only those terms $X^i A_i(Y)$ with $i \leq T$ of the polynomial $\beta_{t}(\boldsymbol{q})$ for $0 \leq t \leq L+m$.


From Corollary~\ref{Corollary3}, the upper bounds may be improved by increasing the truncating parameter $T$, which usually needs more computational and memory loads. However, the lower bound~(as well as the upper bound in the high SNR region) is dominated by the CRWEFs with input weights 1 and 2, which can be given explicitly as below.

We first show the CRWEFs for a systematic BMST-R code ensemble without puncturing. We have
\begin{equation}\label{eq:A1}
    A_1 \left( Y \right) = L\Layerk Y^{(m+1)(N-1)}.
\end{equation}
For the CRWEF $A_2(Y)$, we consider the following three cases.
\begin{enumerate}
  \item The two non-zero information bits are in the same layer. In this case, the corresponding CRWEF $A_2^{(1)} \left( Y \right)$ is given by
      \begin{eqnarray}\label{eq:A2_1}
            A_2^{(1)} \left( Y \right) = \frac{\Layerk(\Layerk-1)L}{2} Y^{2(m+1)(N-1)}.
      \end{eqnarray}

  \item The two non-zero information bits are in two different layers with a gap $\ell$~(spaced away from $\ell-1$ layers) satisfying that $1\leq \ell \leq m$. In this case, the corresponding CRWEF $A_2^{(2)} \left( Y \right)$ is given by
      \ifCLASSOPTIONonecolumn
      \begin{equation}\label{eq:A2_2}
        A_2^{(2)} \left( Y \right) =
        \sum\limits_{\ell=1}^{m} (L-\ell)\Layerk^2
        Y^{2\ell(N-1)}\left(\frac{1}{\Layerk}+\frac{\Layerk-1}{\Layerk}Y^2\right)^{(m+1-\ell)(N-1)}.
      \end{equation}
      \fi
      \ifCLASSOPTIONtwocolumn
      \begin{align}\label{eq:A2_2}
        A_2^{(2)} \left( Y \right) = ~~~~~~~~~~~~~~~~~~~~~~~~~~~~~~~~~~~~~~~~~~~~~~~~~~~~\nonumber \\
        \sum\limits_{\ell=1}^{m} (L-\ell)\Layerk^2
        Y^{2\ell(N-1)}\left(\frac{1}{\Layerk}+\frac{\Layerk-1}{\Layerk}Y^2\right)^{(m+1-\ell)(N-1)}.
      \end{align}
      \fi

  \item The two non-zero information bits are in two different layers with a gap $\ell$ satisfying that $m+1\leq \ell \leq L-1$. In this case, the corresponding CRWEF $A_2^{(3)} \left( Y \right)$ is given by
      \begin{equation}\label{eq:A2_3}
        A_2^{(3)} \left( Y \right) = \sum\limits_{\ell=m+1}^{L-1} (L-\ell)\Layerk^2 \cdot Y^{2(m+1)(N-1)}.
      \end{equation}
\end{enumerate}
In summary, the CRWEF $A_2(Y)$ for a systematic BMST-R code ensemble without puncturing is given by
\begin{equation}\label{eq:A2}
    A_2 \left( Y \right) = A_2^{(1)} \left( Y \right) + A_2^{(2)} \left( Y \right) + A_2^{(3)} \left( Y \right).
\end{equation}

Then, we consider the CRWEFs for a systematic BMST-R code ensemble with $\Layerk_p$ bits in each layer punctured. Taking into account the puncturing effect, when $\Layerk_p$ bits of a sequence with length $\Layerk$ and weight 1 are randomly punctured, the resulting weight enumerator $B_1 \left( Y \right)$ is given by
\begin{eqnarray}\label{eq:B_Y1}
    B_1 \left( Y \right) &=& \frac{\binom{\Layerk-1}{\Layerk_p-1}}{\binom{\Layerk}{\Layerk_p}}
    +\frac{\binom{\Layerk-1}{\Layerk_p}}{\binom{\Layerk}{\Layerk_p}}Y  \nonumber \\
    &=& \frac{\Layerk_p}{\Layerk}+\frac{\Layerk-\Layerk_p}{\Layerk}Y \nonumber \\
    &=& \theta+(1-\theta)Y.
\end{eqnarray}
When $\Layerk_p$ bits of a sequence with length $\Layerk$ and weight 2 are punctured, the resulting weight enumerator $B_2 \left( Y \right)$ is given by
      \ifCLASSOPTIONonecolumn
      \begin{equation}\label{eq:B_Y2}
        B_2 \left( Y \right) =
        \left\{ \begin{array}{cc}
            \frac{2\Layerk_p}{\Layerk}Y + \frac{\Layerk-2\Layerk_p}{\Layerk}Y^2, &\Layerk_p = 0,1,\\
            \frac{\binom{2}{2}\binom{\Layerk-2}{\Layerk_p-2}}{\binom{\Layerk}{\Layerk_p}} + \frac{\binom{2}{1}\binom{\Layerk-2}{\Layerk_p-1}}{\binom{\Layerk}{\Layerk_p}}Y + \frac{\binom{\Layerk-2}{\Layerk_p}}{\binom{\Layerk}{\Layerk_p}}Y^2, &\Layerk_p \geq 2.
        \end{array}\right.
      \end{equation}
      \fi
      \ifCLASSOPTIONtwocolumn
      \begin{align}\label{eq:B_Y2}
        B_2 \left( Y \right) = ~~~~~~~~~~~~~~~~~~~~~~~~~~~~~~~~~~~~~~~~~~~~~~~~~~~~\nonumber\\
        \left\{ \begin{array}{cc}
            \frac{2\Layerk_p}{\Layerk}Y + \frac{\Layerk-2\Layerk_p}{\Layerk}Y^2, &\Layerk_p = 0,1,\\
            \frac{\binom{2}{2}\binom{\Layerk-2}{\Layerk_p-2}}{\binom{\Layerk}{\Layerk_p}} + \frac{\binom{2}{1}\binom{\Layerk-2}{\Layerk_p-1}}{\binom{\Layerk}{\Layerk_p}}Y + \frac{\binom{\Layerk-2}{\Layerk_p}}{\binom{\Layerk}{\Layerk_p}}Y^2, &\Layerk_p \geq 2.
        \end{array}\right.
      \end{align}
      \fi
Then we have
    \ifCLASSOPTIONonecolumn
    \begin{eqnarray}\label{eq:A1-Pun}
        A_1 \left( Y \right) &=& L\Layerk Y^{(m+1)(N-2)}\left(B_1 \left( Y \right)\right)^{m+1} \nonumber \\
        &=& L\Layerk Y^{(m+1)(N-2)} \sum\limits_{\ell=0}^{m+1} \binom{\Memory+1}{\ell}\theta^{m+1-\ell}(1-\theta)^{\ell}Y^{\ell} .
    \end{eqnarray}
    \fi
    \ifCLASSOPTIONtwocolumn
    \begin{eqnarray}\label{eq:A1-Pun}
        A_1 \left( Y \right) = L\Layerk Y^{(m+1)(N-2)}\left(B_1 \left( Y \right)\right)^{m+1} ~~~~~~~~~~~~~~~~\nonumber \\
        = L\Layerk Y^{(m+1)(N-2)} \sum\limits_{\ell=0}^{m+1} \binom{\Memory+1}{\ell}\theta^{m+1-\ell}(1-\theta)^{\ell}Y^{\ell} .
    \end{eqnarray}
    \fi

For the CRWEF $A_2(Y)$, we consider the following three cases.
\begin{enumerate}
  \item The two non-zero information bits are in the same layer. In this case, the corresponding CRWEF $A_2^{(1)} \left( Y \right)$ is given by
      \begin{eqnarray}\label{eq:A2_1-Pun}
            A_2^{(1)} \left( Y \right) = \frac{\Layerk(\Layerk-1)L}{2}Y^{2(m+1)(N-2)}\left( B_2(Y) \right)^{m+1}.
      \end{eqnarray}

  \item The two non-zero information bits are in two different layers with a gap $\ell$ satisfying that $1\leq \ell \leq m$. In this case, the corresponding CRWEF $A_2^{(2)} \left( Y \right)$ is given by
      \begin{align}\label{eq:A2_2-Pun}
        A_2^{(2)} \left( Y \right) = \sum\limits_{\ell=1}^{m} (L-\ell)\Layerk^2
        Y^{2\ell(N-2)} (B_1(Y))^{2\ell} ~~~~~~~~~~~~~\nonumber\\
        \cdot \left(\frac{1}{\Layerk}\!+\!\frac{\Layerk-1}{\Layerk}Y^2\right)^{(m\!+\!1\!-\!\ell)(N\!-\!2)}\!
        \left(\frac{1}{\Layerk}\!+\!\frac{\Layerk\!-\!1}{\Layerk}B_2(Y)\right)^{m\!+\!1\!-\!\ell}.
      \end{align}

  \item The two non-zero information bits are in two different layers with a gap $\ell$ satisfying that $m+1\leq \ell \leq L-1$. In this case, the corresponding CRWEF $A_2^{(3)} \left( Y \right)$ is given by
      \ifCLASSOPTIONonecolumn
      \begin{align}\label{eq:A2_3-Pun}
        A_2^{(3)} \left( Y \right) =
        \sum\limits_{\ell=m+1}^{L-1} (L-\ell)\Layerk^2
        \cdot Y^{2(m+1)(N-2)}\left(B_1(Y)\right)^{2(m+1)}.
      \end{align}
      \fi
      \ifCLASSOPTIONtwocolumn
      \begin{align}\label{eq:A2_3-Pun}
        A_2^{(3)} \left( Y \right) =  ~~~~~~~~~~~~~~~~~~~~~~~~~~~~~~~~~~~~~~~~~~~\nonumber\\
        \sum\limits_{\ell=m+1}^{L-1} (L-\ell)\Layerk^2
        \cdot Y^{2(m+1)(N-2)}\left(B_1(Y)\right)^{2(m+1)}.
      \end{align}
      \fi
\end{enumerate}
In summary, the CRWEF $A_2(Y)$ for a systematic BMST-R code ensemble with $\Layerk_p$ bits in each layer punctured is given by
\begin{equation}\label{eq:A2-Pun}
    A_2 \left( Y \right) = A_2^{(1)} \left( Y \right) + A_2^{(2)} \left( Y \right) + A_2^{(3)} \left( Y \right).
\end{equation}


From Theorem~\ref{MAPLowerBound}, we know that the bit-error probability for systematic codes under MAP decoding can be lower-bounded in terms of the minimum Hamming weights $d_{\min,i}$ of $\mathcal{C}_{1,i}$. However, these minimum weights are usually not available for a general code. If this is the case, we can use instead the row weights of the generator matrix to calculate a looser lower bound as shown in Corollary~\ref{LDGM_LowerBound}, where the $i$-th row weight can be determined by setting the binary information data $\boldsymbol{u}$ to a nonzero sequence with only the $i$-th component $u_i=1$. Then we have the following two corollaries.


\vspace{0.15cm}
\begin{corollary}\label{CorollaryEnsembleLowerBound}
The bit-error probability of a systematic BMST-R code ensemble under MAP decoding can be lower-bounded by
    \ifCLASSOPTIONonecolumn
    \begin{equation}\label{EnsembleLowerBound}
        {\rm BER_{MAP}} \geq \sum_{\ell = 0}^{\Memory+1}\binom{\Memory+1}{\ell}\theta^{m+1-\ell}(1-\theta)^{\ell}
        Q\left(\frac{\sqrt{N + \Memory(N-2)-1 + \ell}}{\sigma}\right),
    \end{equation}
    \fi
    \ifCLASSOPTIONtwocolumn
    \begin{align}\label{EnsembleLowerBound}
        {\rm BER_{MAP}} \geq \sum_{\ell = 0}^{\Memory+1}\binom{\Memory+1}{\ell}\theta^{m+1-\ell}(1-\theta)^{\ell} ~~~~~~~~~~~~\nonumber\\
        Q\left(\frac{\sqrt{N + \Memory(N-2)-1 + \ell}}{\sigma}\right),
    \end{align}
    \fi
where $\theta$ is the puncturing fraction.
\end{corollary}

\begin{IEEEproof}
Due to the random puncturing, the $i$-th row weight $W_i$ of the generator matrix for a systematic BMST-R code ensemble is a random variable. Given a puncturing fraction $\theta$, the probability mass function of $W_i$ can be calculated as
\begin{equation}\label{prob}
    {\rm Pr}\left\{W_i=N + \Memory(N-2)-1 + \ell \right\} = \binom{\Memory+1}{\ell}\theta^{m+1-\ell}(1-\theta)^{\ell},
\end{equation}
where $\ell \in \{0,1,\cdots,\Memory+1\}$. Thus, the error probability of the $i$-th estimated bit of the systematic BMST-R code ensemble under MAP decoding can be lower-bounded by
\ifCLASSOPTIONonecolumn
\begin{eqnarray}\label{BitEnsembleLowerBound}
    {\rm Pr}\{E_i\}_{\rm MAP}
    &\geq& \mathbb{E}\left[ Q \left(\frac{\sqrt{W_i}}{\sigma}\right) \right] \nonumber\\
    &=& \sum_{\ell = 0}^{\Memory+1}\binom{\Memory\!+\!1}{\ell}\theta^{m+1-\ell}(1-\theta)^{\ell} Q\left(\frac{\sqrt{N + \Memory(N-2)-1 + \ell}}{\sigma}\right).
\end{eqnarray}
\fi
\ifCLASSOPTIONtwocolumn
\begin{equation}\label{BitEnsembleLowerBound}
\begin{split}
    {\rm Pr}\{E_i\}_{\rm MAP}
    &\geq \mathbb{E}\left[ Q \left(\frac{\sqrt{W_i}}{\sigma}\right) \right] \\
    &= \sum_{\ell = 0}^{\Memory+1}\binom{\Memory\!+\!1}{\ell}\theta^{m+1-\ell}(1-\theta)^{\ell}\\
    &~~~~~~~~~~~~~~~Q\left(\frac{\sqrt{N + \Memory(N-2)-1 + \ell}}{\sigma}\right).
\end{split}
\end{equation}
\fi
It follows that the bit-error probability of the systematic BMST-R code ensemble under MAP decoding can be lower-bounded by
\ifCLASSOPTIONonecolumn
\begin{eqnarray}
    {\rm BER_{MAP}} &=& \frac{1}{\Allk} \sum\limits_{i=0}^{\Allk-1} {\rm Pr}\{E_i\}_{\rm MAP} \nonumber\\
    &\geq& \sum_{\ell = 0}^{\Memory+1}\binom{\Memory+1}{\ell}\theta^{m+1-\ell}(1-\theta)^{\ell}
    Q\left(\frac{\sqrt{N + \Memory(N-2)-1 + \ell}}{\sigma}\right).
\end{eqnarray}
\fi
\ifCLASSOPTIONtwocolumn
\begin{equation}
\begin{split}
    {\rm BER_{MAP}} &= \frac{1}{\Allk} \sum\limits_{i=0}^{\Allk-1} {\rm Pr}\{E_i\}_{\rm MAP} \\
    &\geq \sum_{\ell = 0}^{\Memory+1}\binom{\Memory+1}{\ell}\theta^{m+1-\ell}(1-\theta)^{\ell}\\
    &~~~~~~~~~~~~~\cdot Q\left(\frac{\sqrt{N + \Memory(N-2)-1 + \ell}}{\sigma}\right).\\
\end{split}
\end{equation}
\fi
\end{IEEEproof}

\vspace{0.15cm}
\begin{corollary}\label{Corollary8}
The bit-error probability of a systematic BMST-R code ensemble without puncturing~(i.e., with puncturing fraction $\theta=0$) under MAP decoding can be lower-bounded by
\begin{equation}\label{CodeLowerBound}
    {\rm BER_{MAP}} \geq \Qfun{\frac{\sqrt{N + \Memory(N-1)}}{\sigma}}.
\end{equation}
\end{corollary}

\begin{IEEEproof}
For a specific $i$~$(0 \leq i \leq \Allk-1)$, we can see from~(\ref{eq:A1}) that the $i$-th row of the generator matrix has a deterministic weight
\begin{equation}\label{d_i}
  w_i = N + \Memory(N-1).
\end{equation}
Thus, the error probability of the $i$-th estimated bit of the systematic BMST-R code ensemble under MAP decoding can be lower-bounded by
\begin{eqnarray}
    {\rm Pr}\{E_i\}_{\rm MAP}
    &\geq& Q \left(\frac{\sqrt{w_i}}{\sigma}\right) \nonumber\\
    &=& Q \left(\frac{\sqrt{N + \Memory(N-1)}}{\sigma} \right).
\end{eqnarray}
It follows that the bit-error probability of the systematic BMST-R code ensemble without puncturing under MAP decoding can be lower-bounded by
\begin{eqnarray}
    {\rm BER_{MAP}} &=& \frac{1}{\Allk} \sum\limits_{i=0}^{\Allk-1} {\rm Pr}\{E_i\}_{\rm MAP} \nonumber\\
    &\geq& Q \left(\frac{\sqrt{N + \Memory(N-1)}}{\sigma} \right).
\end{eqnarray}
\end{IEEEproof}

\textbf{Remarks.}~Corollaries~\ref{CorollaryEnsembleLowerBound} and~\ref{Corollary8} also hold for systematic BMST-R codes with specific interleavers for the reason that the interleavers have no effect on the row weights of the generator matrix. Since the lower bound~(\ref{CodeLowerBound}) without puncturing is equivalent to the lower bound~(\ref{EnsembleLowerBound}) with puncturing fraction $\theta=0$, in the rest of the paper, we consider for generality the lower bound~(\ref{EnsembleLowerBound}).

\subsection{Trade-Off Between Performance and Complexity}\label{subsec:TradeOff}

\begin{figure}[t]
    \center
    \includegraphics[clip, width=\figwidth]{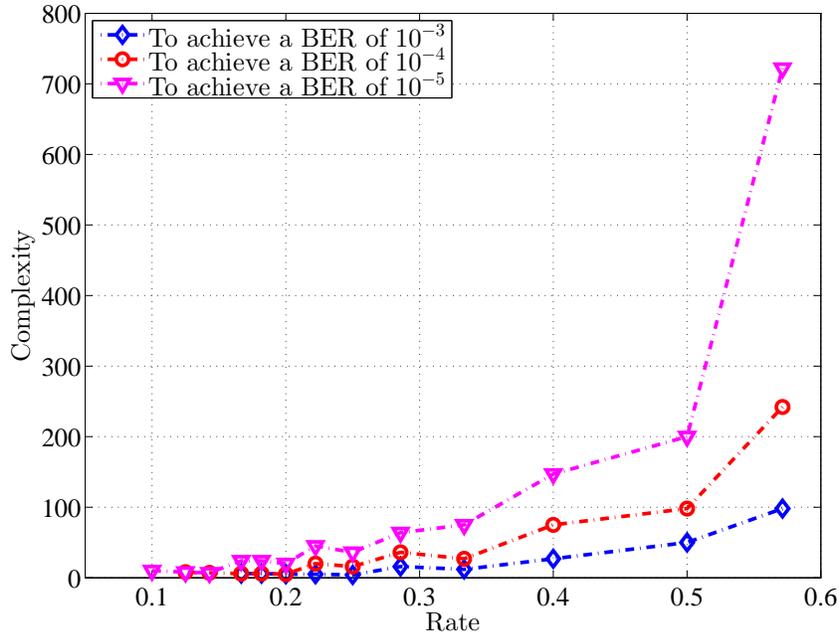}%
    \caption{Decoding complexity for systematic BMST-R codes as a function of code rate that requires an SNR of 2 dB to achieve target BERs of $10^{-3}$, $10^{-4}$ and $10^{-5}$.}
    \label{Fig_Complexity_DiffRate}
\end{figure}

\begin{figure}[t]
    \center
    \includegraphics[clip, width=\figwidth]{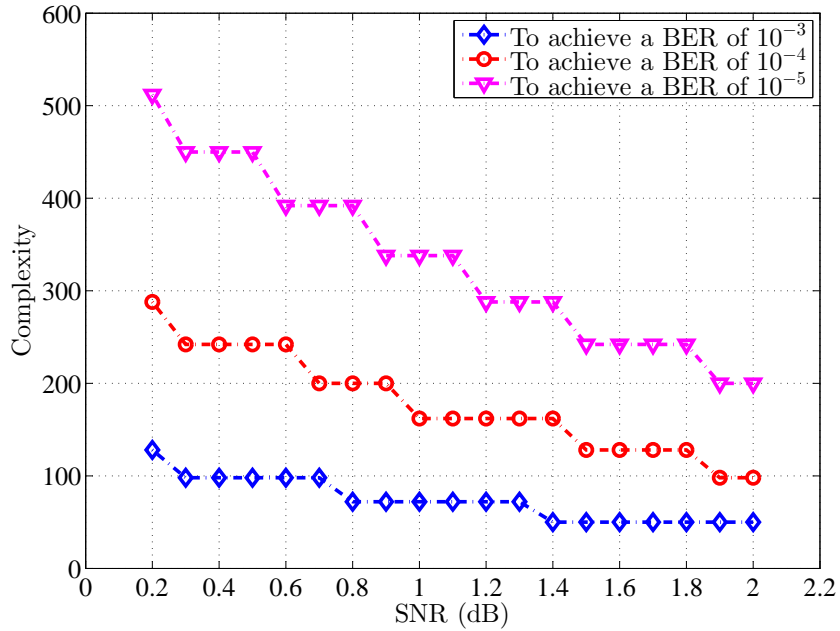}%
    \caption{Decoding complexity for rate 1/2 systematic BMST-R codes as a function of SNR with target BERs of $10^{-3}$, $10^{-4}$ and $10^{-5}$.}
    \label{Fig_Complexity_Rate0.5}
\end{figure}

The implementation complexity of systematic BMST-R codes can be analyzed as with their non-systematic counterpart. For encoding, the information sequence is partitioned equally into blocks and transmitted directly, while their replicas are interleaved and transmitted in a block Markov superposition manner. This shows that the encoding complexity grows linearly with the encoding memory $\Memory$. For decoding, a sliding window decoding algorithm with a tunable decoding delay can be implemented over the normal graph~(see Fig.~\ref{BMST_decoder}). The decoding complexity for node \fbox{+} and node \fbox{=} of systematic BMST-R codes grows linearly with the encoding memory $\Memory$. Furthermore, similar to non-systematic BMST codes, a decoding delay $d = 2m \!\sim\! 3 m$ is required to achieve the lower bound on the performance. As a result, the decoding complexity for systematic BMST-R codes can be roughly given as $\mathcal{O}(N\Memory d)$, or equivalently, $\mathcal{O}(N\Memory^2)$.


The above analysis shows that the decoding complexity is closely related to the repetition degree $N$ and the encoding memory $\Memory$, both of which in turn determine the lower bound~(\ref{EnsembleLowerBound}). This allows us to make trade-offs among efficiency, performance and complexity. To be precise, we consider the following two cases.
\begin{enumerate}
  \item Fixed SNR. We can observe from the lower bound~(\ref{EnsembleLowerBound}) that, for a given SNR and BER, the required encoding memory $\Memory$ decreases as the repetition degree $N$ increases~(accordingly, the code rate decreases), resulting in a lower complexity. Fig.~\ref{Fig_Complexity_DiffRate} shows the decoding complexity for systematic BMST-R codes as a function of code rate that requires an SNR of 2 dB to achieve BERs of $10^{-3}$, $10^{-4}$ and $10^{-5}$. As expected, for fixed BER, the greater the code rate is, the higher the decoding complexity is. We also see that for fixed code rate, the higher the performance requirement~(equivalently, the more stringent the BER) is, the higher the decoding complexity is.
  \item Fixed code rate. We can observe from the lower bound~(\ref{EnsembleLowerBound}) that, for a given rate and BER, the required encoding memory $\Memory$ decreases as the SNR increases, resulting in a lower decoding complexity. This is reasonable since more excessive SNR is available compared to the corresponding Shannon limit. Fig.~\ref{Fig_Complexity_Rate0.5} shows the decoding complexity for rate 1/2 systematic BMST-R codes as a function of SNR with BERs of $10^{-3}$, $10^{-4}$ and $10^{-5}$. As expected, for fixed BER, the greater the SNR is, the lower the decoding complexity is. We also observe that for fixed SNR, the more stringent the BER is, the higher the decoding complexity is.
\end{enumerate}

\section{Numerical Analysis and Performance Comparison}\label{SecIV}
In this section, all simulations are performed assuming BPSK modulation and transmitted over an AWGN channel, unless otherwise specified. The $(m+1)(N-1)$ random interleavers~(randomly generated but fixed) of size $\Layerk$ are used for encoding. The iterative sliding window decoding algorithm for systematic BMST-R codes is performed using the parallel (flooding) updating schedule within the decoding window with a maximum iteration number of 18, and the entropy stopping criterion~\cite{Ma04,Ma15} with a preselected threshold of $10^{-6}$ is employed.

\subsection{Performance Bounds and Code Construction}\label{SecIV-A}

In this subsection, we present an example to study the performance bounds on BER of systematic BMST-R codes. We consider systematic BMST-R codes with repetition degree $N=2$ and puncturing fraction $\theta=0$. The decoding delay $d$ for the sliding window decoding is specified as $d=3m$.

\begin{example}\label{Example1}
\begin{figure}[t]
    \center
    \includegraphics[clip, width=\figwidth]{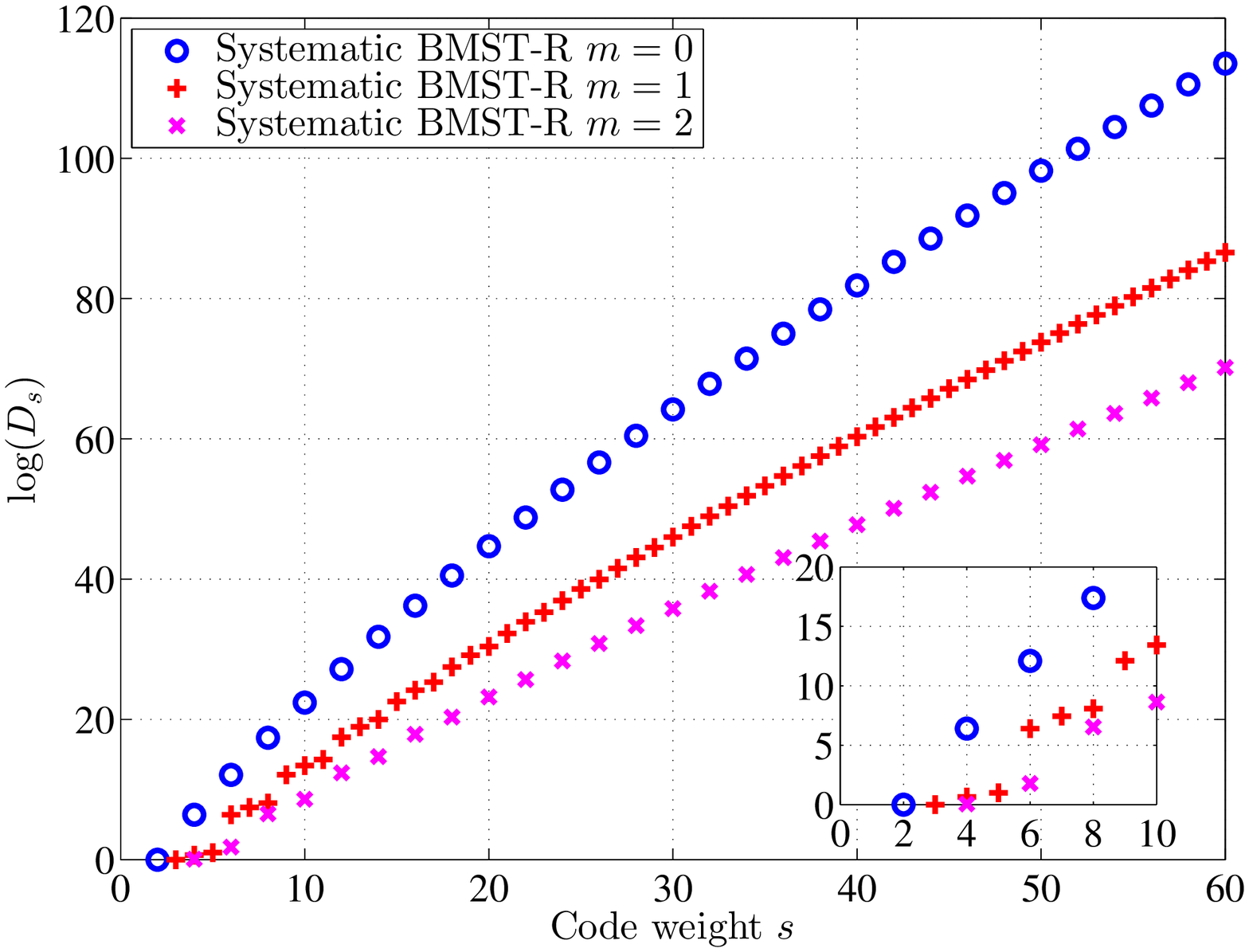}%
    \caption{Spectrum $\{D_s\}~(0\leq s\leq 60)$ of systematic BMST-R code ensembles with encoding encoding memory $m=0$, $m=1$ and $m=2$ in Example 1. Assume $L=20$ blocks of information data, where the information subsequence has length $\Layerk=30$. The systematic BMST-R code with $m=0$ is equivalent to the independent transmission of rate 0.5 repetition code. The truncating parameter is set to $T=60$. The code rates of systematic BMST-R code ensembles with $m=0$, $m=1$ and $m=2$ are 0.5, 0.4878 and 0.4762, respectively.}
    \label{Fig_Spectrum}
\end{figure}
\begin{figure}[t]
    \center
    \includegraphics[clip, width=\figwidth]{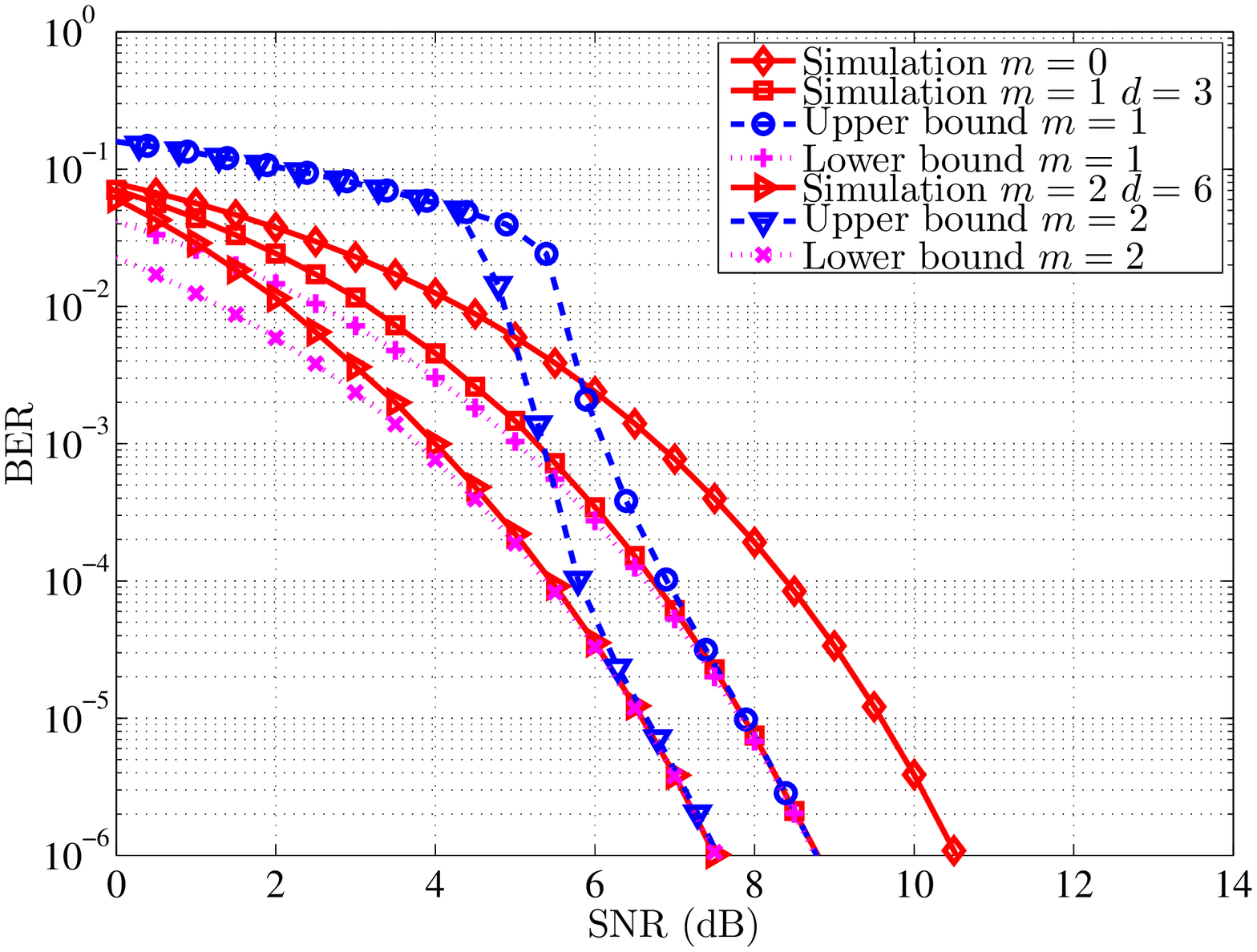}%
    \caption{Performance of systematic BMST-R code ensembles with encoding encoding memory $m=0$, $m=1$ and $m=2$ in Example 1. The systematic BMST-R code with $m=0$ is equivalent to the independent transmission of rate 0.5 repetition code. Assume $L=20$ blocks of information data, where the information subsequence has length $\Layerk=30$. The codeword is modulated using BPSK and transmitted over an AWGN channel. The decoding delay $d$ is specified as $d=3m$. The truncating parameter is set to $T=60$. The code rates of systematic BMST-R code ensembles with $m=0$, $m=1$ and $m=2$ are 0.5, 0.4878 and 0.4762, respectively.}
    \label{Fig_PerformanceBounds}
\end{figure}
Assume that there are $L=20$ blocks of information data to be transmitted, where the information subsequence have length $\Layerk=30$. We consider systematic BMST-R code ensembles with encoding memory $m=0$, $m=1$ and $m=2$, whose code rates are 0.5, 0.4878 and 0.4762, respectively. Here, the systematic BMST-R code with $m=0$ is equivalent to the independent transmission of rate 0.5 repetition code. Assume that we only calculate the truncated IRWEF $\{A_{i,j}$, $0 \leq i \leq 60\}$. That is, the truncating parameter is set to $T=60$. Fig.~\ref{Fig_Spectrum} shows the spectrum $\{D_s\}~(0< s\leq T)$ of systematic BMST-R code ensembles, where
\begin{equation}\label{eq:Ds}
  D_{s} = \sum_{i=1}^{T}\frac{i}{\Allk}A_{i,s-i}.
\end{equation}
From Fig.~\ref{Fig_Spectrum}, we see that the spectrum of the systematic BMST-R code ensembles with $m=1$ and $m=2$ have less number of codewords with small Hamming weights. We also see that the minimum Hamming distances of systematic BMST-R code ensembles with encoding memory $m=0$, $m=1$ and $m=2$ are 2, 3 and 4, respectively. These indicate that the systematic BMST-R codes have potentially better performance than the independent transmission system. The simulation results are shown in Fig.~\ref{Fig_PerformanceBounds}, where we observe that
\begin{enumerate}
  \item The lower and upper bounds on the BER performance of systematic BMST-R codes are tight in the high SNR region.
  \item The simulated BER performance curves match well with the bounds in the high SNR region, indicating that the sliding window decoding algorithm is near optimal in the high SNR region.
  \item The systematic BMST-R codes outperform the independent transmission system~(i.e., the systematic BMST-R code with $m=0$). Furthermore, the systematic BMST-R code with encoding memory $m=2$ outperforms the systematic BMST-R code with $m=1$. Taking into account the rate loss, the systematic BMST-R code with $m=2$ obtains a net gain of 1.175 dB in terms of $E_b/N_0$ at a BER of $10^{-5}$, compared to the systematic BMST-R code with $m=1$.
\end{enumerate}
\end{example}

Given the tightness of the lower bound~(\ref{EnsembleLowerBound}) as demonstrated in Example~\ref{Example1}, we have the following simple procedure to construct good codes. Let $R \in (0,1)$ be the target code rate and $p_{\rm target}$ be the target BER. The object is to construct a code with rate $R_L \approx R$, which can approach the Shannon limit at the target BER. A systematic BMST-R code has the following five parameters: repetition degree $N$, information subsequence length $\Layerk$, puncturing length $\Layerk_p$, data block length $L$, and encoding memory $m$. These parameters can be determined as follows.


\begin{enumerate}
  \item Determine the repetition degree $N$ and puncturing fraction $\theta$ such that $\frac{1}{N-\theta} = R$. Choose sufficiently large information subsequence length\footnote{By simulation, we find that $\Layerk\geq 2500$ suffices to approach the Shannon limit within around half dB.} $\Layerk$ and puncturing length $\Layerk_p$ such that $\Layerk_p/\Layerk \approx \theta$;
  \item Find the Shannon limit for the given code rate $R$ and target BER $p_{\rm target}$;
  \item Determine the minimum encoding memory $m$ such that the lower bound of ${\rm BER_{MAP}}$ in~(\ref{EnsembleLowerBound}) at the Shannon limit is not greater than the preselected target BER $p_{\rm target}$;
  \item Choose a data block length $L$ such that the rate loss (i.e., $R-R_L$) due to the termination is small;
  \item Generate $(m+1)(N-1)$ interleavers randomly.
\end{enumerate}

Evidently, the above procedure requires no optimization and hence can be easily implemented. The encoding memories for some systematic BMST-R codes required to approach the corresponding Shannon limits at given target BERs are shown in Table~\ref{Table1}. As expected, the lower the target BER is, the greater the required encoding memory $\Memory$ is.

\begin{table*}
\caption{Encoding memories for systematic BMST-R codes required to approach the corresponding Shannon limits at given target BERs}\label{Table1}
  \centering
  \begin{tabular}{|c||c|c|c|c|}
  \hline
  \multirow{2}{*}{Systematic BMST-R Codes} &\multicolumn{4}{c|}{Encoding Memory $m$}\\ \cline{2-5}
   &${\rm BER}=10^{-3}$ &${\rm BER}=10^{-4}$ &${\rm BER}=10^{-5}$ &${\rm BER}=10^{-6}$\\ \hline
  ${\rm Rate}~2/3$, $N=2$, $\theta=0.5$                    &12 &18 &24 &31 \\\hline
  ${\rm Rate}~1/2$, $N=2$, $\theta=0$                       &8 &12 &16 &20  \\\hline
  ${\rm Rate}~2/5$, $N=3$, $\theta=0.5$                     &8 &11 &15 &19  \\\hline
  ${\rm Rate}~1/3$, $N=3$, $\theta=0$                      &7 &11 &14 &18 \\\hline
  ${\rm Rate}~1/4$, $N=4$, $\theta=0$                       &7 &10 &14 &17  \\\hline
\end{tabular}
\end{table*}

\subsection{Impact of Parameters on BER}
In this subsection, we study the impact of various parameters~(e.g., encoding memory $\Memory$, information subsequence length $\Layerk$ and decoding delay $d$) on the BER performance of systematic BMST-R codes with fixed code rate. Note that, as pointed out in~Section~\ref{subsec:encoding}, varying repetition degree $N$ and puncturing fraction $\theta$ results in systematic codes with different rates. For simplicity, we focus on $R_L=0.49$ systematic BMST-R codes with repetition degree $N=2$ and puncturing fraction $\theta=0$. Three regimes are considered: (1)~fixed $\Memory$ and $\Layerk$, increasing $d$, (2)~fixed $\Memory$ and $d$, increasing $\Layerk$, and (3)~fixed $\Layerk$, increasing $\Memory$~(and hence $d$).

\begin{example}[Fixed $\Memory$ and $\Layerk$, Increasing $d$]
\begin{figure}[t]
    \center
    \includegraphics[clip, width=\figwidth]{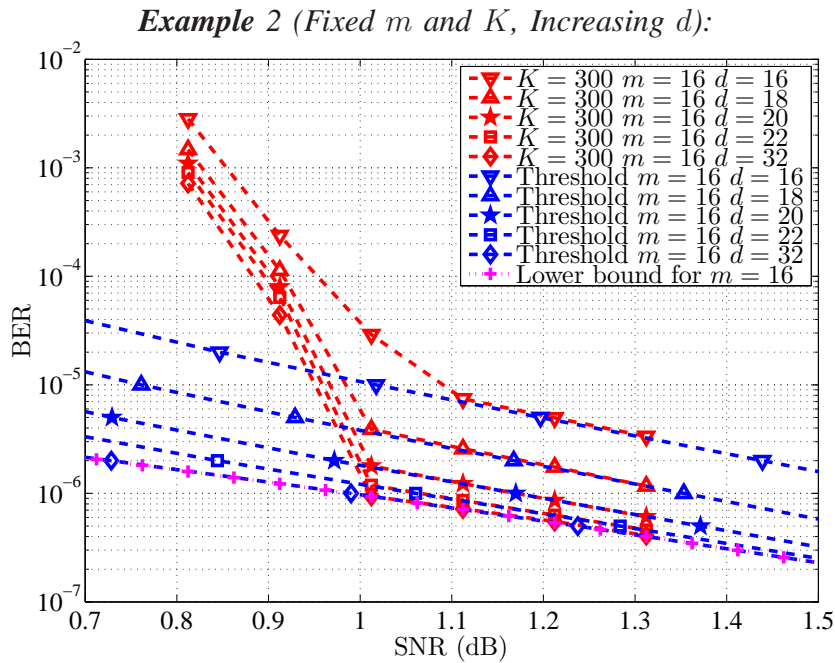}%
    \caption{Simulated decoding performance of rate $R_L=0.49$ systematic BMST-R codes decoded with different decoding delays $d$ in Example 2. Information subsequence length $\Layerk=300$, encoding memory $\Memory=16$ and data block length $L=392$. The codeword is modulated using BPSK and transmitted over an AWGN channel. The corresponding window decoding thresholds and the lower bound are also plotted.}
    \label{Fig_ImpactParameters_DiffDelay}
\end{figure}
Consider a systematic BMST-R code with information subsequence length $\Layerk=300$, encoding memory $\Memory=16$ and data block length $L=392$. The BER performance of the systematic BMST-R code decoded with different decoding delays $d$ is shown in Fig.~\ref{Fig_ImpactParameters_DiffDelay}. The asymptotic threshold performance obtained by using the EXIT chart analysis method in~\cite{Huang15JSAC} is also included. From Fig.~\ref{Fig_ImpactParameters_DiffDelay}, we observe that
\begin{enumerate}
  \item The BER performance of the systematic BMST-R code decoded with different delays $d$ matches well with the corresponding window decoding thresholds in the high SNR region.
  \item The BER performance in the waterfall region improves as the decoding delay $d$ increases, but it does not improve much further beyond a certain decoding delay~(roughly $d=20$).
  \item The BER performance in the error floor region improves as the decoding delay $d$ increases, and it matches well with the lower bound for the systematic BMST-R code with $\Memory=16$ when $d$ increases up to a certain point~(roughly $d=32$).
\end{enumerate}

\end{example}

\begin{example}[Fixed $\Memory$ and $d$, Increasing $\Layerk$]
\begin{figure}[t]
    \center
    \includegraphics[clip, width=\figwidth]{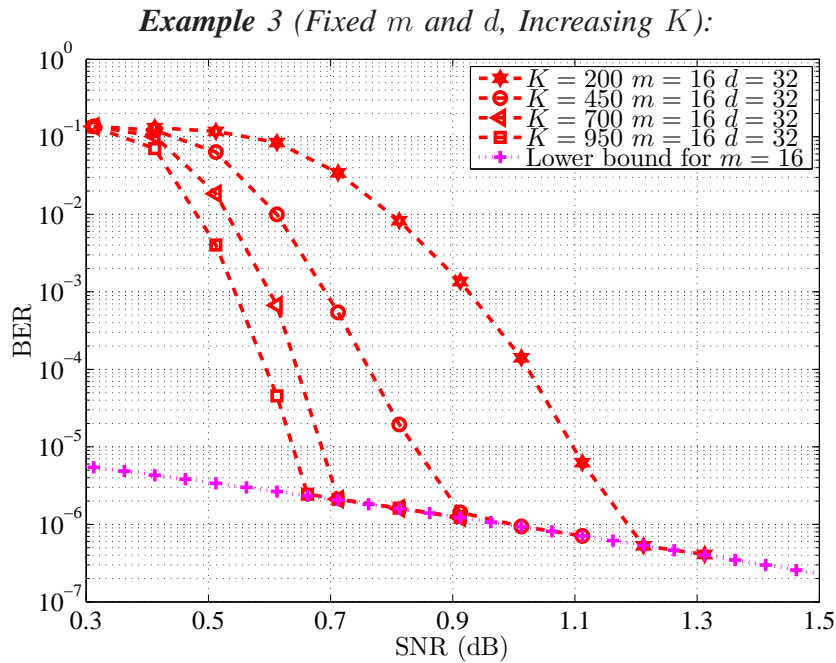}%
    \caption{Simulated decoding performance of rate $R_L=0.49$ systematic BMST-R codes with different information subsequence lengths $\Layerk$ in Example 3. The codes are constructed with encoding memory $\Memory=16$ and data block length $L=392$, and decoded with decoding delay $d=32$. The codeword is modulated using BPSK and transmitted over an AWGN channel. The corresponding lower bound is also plotted.}
    \label{Fig_ImpactParameters_DiffCartesianProductOrder}
\end{figure}
Consider systematic BMST-R codes constructed with encoding memory $\Memory=16$, data block length $L=392$ and decoded with decoding delay $d=32$. The BER performance of systematic BMST-R codes constructed with different information subsequence lengths $\Layerk$ is shown in Fig.~\ref{Fig_ImpactParameters_DiffCartesianProductOrder}, where we observe that
\begin{enumerate}
  \item Increasing the information subsequence length $\Layerk$ can improve waterfall region performance. As expected, this improvement saturates for sufficiently large $\Layerk$. For example, the improvement at a BER of $10^{-5}$ from $\Layerk=200$ to $\Layerk=450$, both decoded with $d=16$, is about 0.25 dB, while the improvement decreases to about 0.05 dB from $\Layerk=700$ to $\Layerk=950$.

  \item The error floors, which are determined by the encoding memory and code rate~(see Corollary~\ref{CorollaryEnsembleLowerBound}), cannot be lowered by increasing $\Layerk$.
\end{enumerate}
\end{example}

\begin{example}[Fixed $\Layerk$, Increasing $\Memory$~(and hence $d$)]
\begin{figure}[t]
  \centering
  \includegraphics[width=\figwidth]{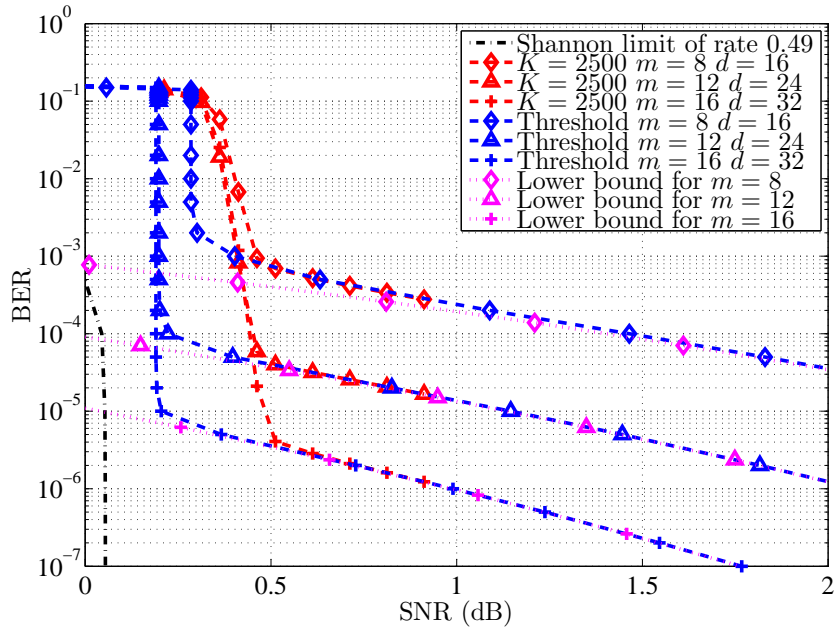}
  \caption{Simulated decoding performance of rate $R_L=0.49$ systematic BMST-R codes constructed with different encoding memories $\Memory$ and decoded with decoding delay $d=2\Memory$ in Example 4. The information subsequence length of the involved systematic BMST-R codes is $\Layerk=2500$. The codeword is modulated using BPSK and transmitted over an AWGN channel. The corresponding window decoding thresholds and lower bounds for systematic BMST-R codes are also plotted.}
  \label{Fig_ImpactParameters_DiffMemory}
\end{figure}
Consider systematic BMST-R codes constructed with information subsequence length $\Layerk=2500$ and encoding memories $\Memory=8$, $12$ and $16$. The performance with sufficiently large decoding delay $d=2\Memory$ of the systematic BMST-R codes is shown in Fig.~\ref{Fig_ImpactParameters_DiffMemory}, where we observe that
\begin{enumerate}
  \item The BER performance of systematic BMST-R codes matches well with the corresponding window decoding thresholds in the high SNR region.
  \item For a high target BER~(roughly above $10^{-3}$), the BER performance improves as the encoding memory $\Memory$ increases, which is consistent with the threshold analysis performance. Note that this phenomenon does not exist for non-systematic BMST codes whose performance degrades slightly as $\Memory$ increases~(see Section~V-C of~\cite{Huang15JSAC}).
  \item The error floor can be lowered by increasing the encoding memory $\Memory$~(and hence the decoding delay $d$).
\end{enumerate}
\end{example}

\subsection{Decoding Performance Based on Latency}
In addition to decoding performance, the latency introduced by employing channel coding is a crucial factor in the design of a practical communication system, such as personal wireless communication and real-time audio and video. In this subsection, we study the BER performance of systematic BMST-R codes based on their decoding latency.

\begin{figure}[t]
  \centering
  \includegraphics[clip,width=\figwidth]{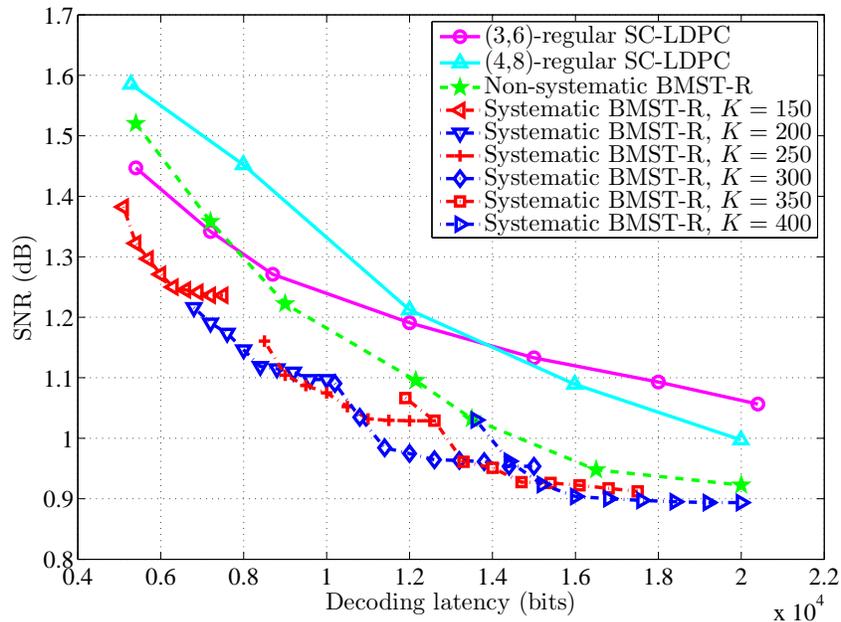}
  \caption{Required SNR to achieve a BER of $10^{-5}$ for finite-length systematic BMST-R codes, non-systematic BMST-R codes in~\cite{Huang15JSAC}, $(3,6)$-regular SC-LDPC codes, and $(4,8)$-regular SC-LDPC codes as a function of decoding latency. All the codes have rate 0.49. The decoding delays for $(3,6)$-regular SC-LDPC codes and $(4,8)$-regular SC-LDPC codes are $5$ and $3$, respectively. The encoding memories for non-systematic BMST-R codes and systematic BMST-R codes are 8 and 16, respectively. The values of the information subsequence length and decoding delay for the non-systematic BMST-R codes are chosen such that the combination gives the best decoding performance. The decoding delays for the systematic BMST-R codes are $d=16$, $17$, $\cdots$, $24$. The codeword is modulated using BPSK and transmitted over an AWGN channel.}
  \label{Fig_EqualDecodingLatency}
\end{figure}

\begin{example}
We consider rate $R_L=0.49$ systematic BMST-R codes with encoding memory $\Memory=16$, repetition degree $N=2$ and puncturing fraction $\theta=0$. The decoding latency of the sliding window decoder, in terms of bits, is given by
\begin{equation}\label{BMST_RC_latency}
    \tau=2\Layerk(d+1).
\end{equation}
The SNR required to achieve a BER of $10^{-5}$ as a function of decoding latency is shown in Fig.~\ref{Fig_EqualDecodingLatency}. We observe that the performance of systematic BMST-R codes~(with fixed information subsequence length $\Layerk$) improves as the decoding delay $d$~(and hence the latency) increases, but it does not improve much further beyond a certain decoding delay. Moreover, beyond a certain latency, using a greater information subsequence length $\Layerk$ with a smaller decoding delay $d$ gives better performance. For example, the systematic BMST-R code constructed with a greater information subsequence length $\Layerk=300$ and decoded with a smaller decoding delay $d=19$ outperforms the systematic BMST-R code constructed with a small information subsequence length $\Layerk=250$ and decoded with a greater decoding delay $d=23$~(both have the same decoding latency of 12000 bits).

We also compare the performance of systematic BMST-R codes, non-systematic BMST-R codes in~\cite{Huang15JSAC}, and SC-LDPC codes when the decoding latencies are equal, as shown in Fig.~\ref{Fig_EqualDecodingLatency}. All the codes have rate 0.49. We restrict consideration to $(3,6)$-regular SC-LDPC codes with two component submatrices $\mathbf{B}_{0}=[2~1]$ and $\mathbf{B}_{1}=[1~2]$, and $(4,8)$-regular SC-LDPC codes with two component submatrices $\mathbf{B}_{0}=[3~1]$ and $\mathbf{B}_{1}=[1~3]$. The decoding delays for $(3,6)$-regular SC-LDPC codes and $(4,8)$-regular SC-LDPC codes are $5$ and $3$, respectively, which are good choices to achieve optimum performance when the decoding latencies are fixed.\footnote{For a more in-depth discussion of the relationship between the protograph lifting factor, the decoding window size and the decoding performance of SC-LDPC codes when the decoding latency is fixed, we refer the reader to~\cite{Huang14}.} The encoding memories for non-systematic BMST-R codes and systematic BMST-R codes are 8 and 16, respectively. The values of the information subsequence length and decoding delay for the non-systematic BMST-R codes are chosen such that the combination gives the best decoding performance~(see Section~VI-A of~\cite{Huang15JSAC}). The decoding delays for the systematic BMST-R codes are $d=16$, $17$, $\cdots$, $24$. We observe that the systematic BMST-R codes perform better than both the non-systematic BMST-R codes and the SC-LDPC codes. For example, in the decoding latency of 12000 bits, the systematic BMST-R code with information subsequence length $\Layerk=300$ and decoding delay $d=19$ gains 0.12 dB, 0.21 dB and 0.24 dB, respectively, compared to the non-systematic BMST-R code, $(3,6)$-regular SC-LDPC code, and $(4,8)$-regular SC-LDPC code.
\end{example}

\subsection{Rate-Compatible Property}
In this subsection, we show the performance of systematic BMST-R codes with different rates by varying repetition degree $N$ and puncturing fraction $\theta$.

\begin{figure}[t]
  \centering
  \includegraphics[width=\figwidth]{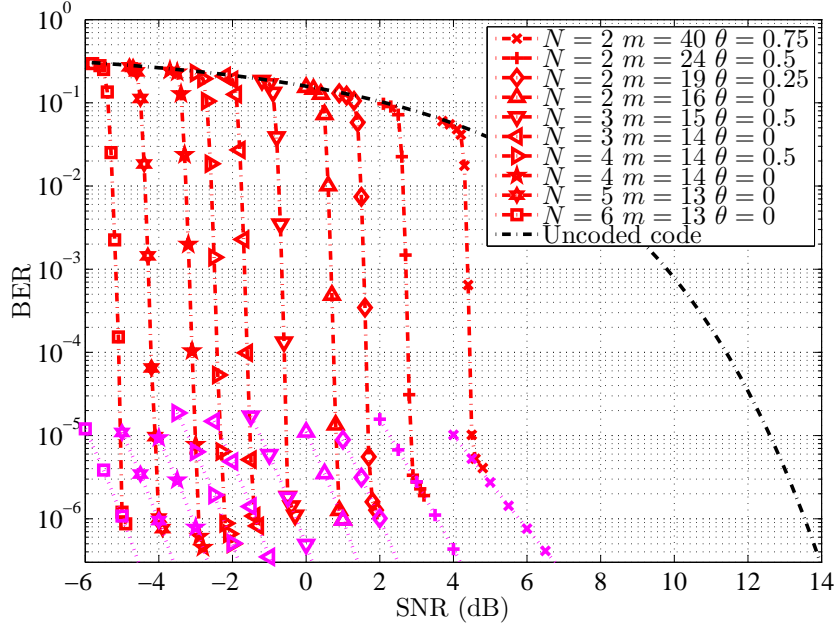}
  \caption{Simulated decoding performance of systematic BMST-R codes with information subsequence length $\Layerk=500$ and data block length $L=500$. The repetition degree $N$, encoding memories $m$ and puncturing fraction $\theta$ are specified in the legends. The decoding delay is specified as $d=2m$. The codeword is modulated using BPSK and transmitted over an AWGN channel. The corresponding lower bounds~(dotted magenta) for systematic BMST-R codes are also plotted. The rates of the systematic BMST-R codes corresponding to the BER curves from left to right in the figure are $0.1631,0.1959,0.2449, 0.2801, 0.3272, 0.3929, 0.4921, 0.5623, 0.6562,$ and $0.7874$.}
  \label{Fig_BERMultiRate}
\end{figure}

\begin{figure}[t]
  \centering
  \includegraphics[width=\figwidth]{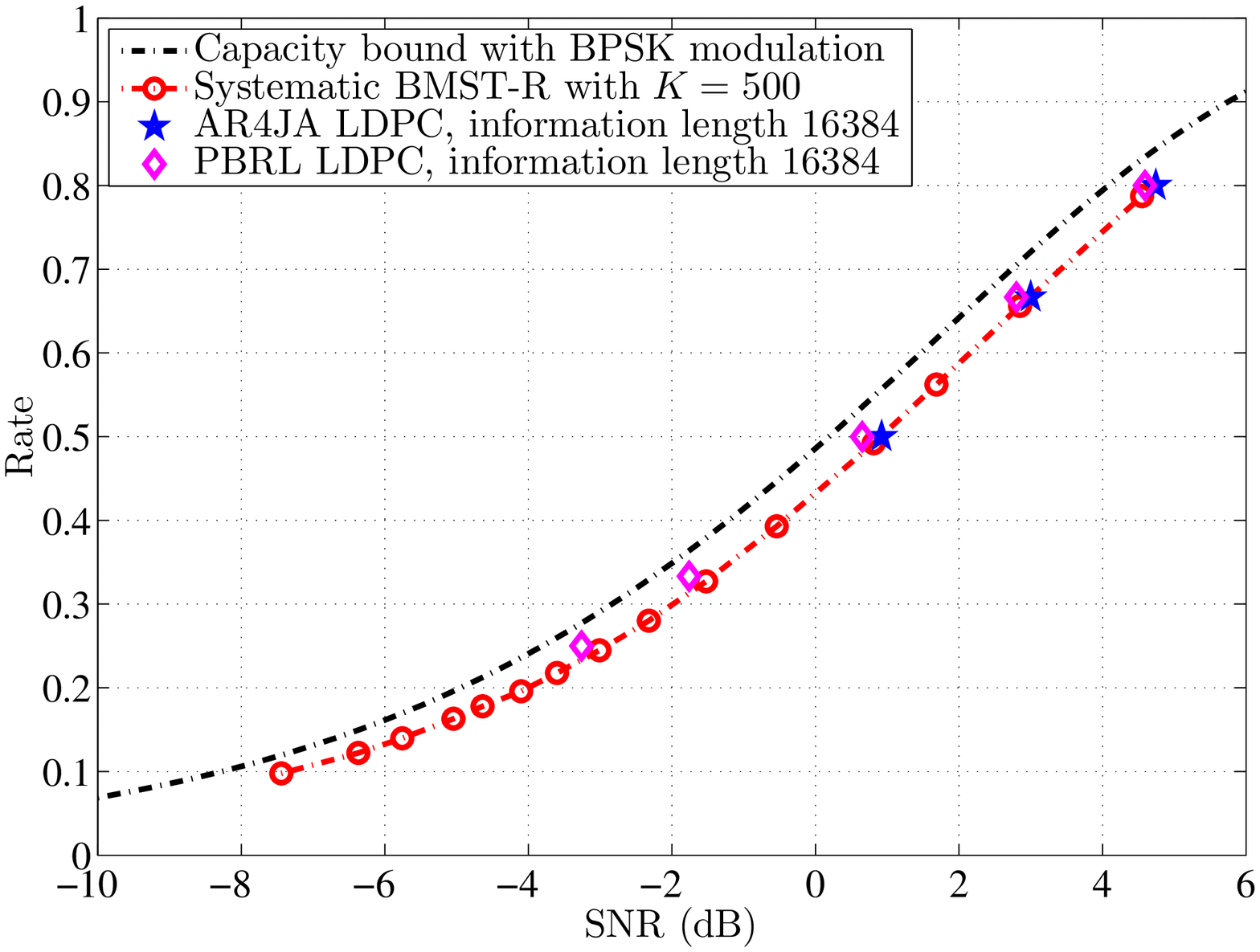}
  \caption{Required SNR to achieve a BER of $10^{-5}$ for systematic BMST-R codes with information subsequence length $\Layerk=500$. The codeword is modulated using BPSK and transmitted over an AWGN channel. The performances of three AR4JA LDPC codes with code rates $1/2$, $2/3$ and $4/5$ in the CCSDS standard~\cite{CCSDS12Coding}, and five PBRL LDPC codes~\cite{Chen15} with code rates $1/4$, $1/3$, $1/2$, $2/3$, and $4/5$, all of which have information length $16384$, are also included.}
  \label{Fig_RequiredSNRMultiRate}
\end{figure}

\begin{example}
Consider systematic BMST-R codes with information subsequence length $\Layerk=500$ and data block length $L=500$. The encoding memories $m$ for systematic BMST-R codes required to approach the Shannon limits at a target BER of $10^{-5}$ are determined following the procedure described in Section~\ref{SecIV-A}. The decoding delay is specified as $d=2m$. Simulation results for systematic BMST-R codes with different rates are shown in Fig.~\ref{Fig_BERMultiRate}. We observe that the performances for all code rates are almost the same as that for uncoded code in the relatively low SNR region. This is different from non-systematic BMST codes whose performance in the relatively low SNR region is very bad due to error propagation. We also observe that, as the SNR increases, the performance curves of the systematic BMST-R codes drop down to the respective lower bounds for all considered code rates.

To evaluate the bandwidth efficiency, we plot the required SNR to achieve a BER of $10^{-5}$ of the systematic BMST-R codes with information subsequence length $\Layerk=500$ against the code rate in Fig.~\ref{Fig_RequiredSNRMultiRate}, where we observe that the systematic BMST-R codes achieve the BER of $10^{-5}$ within one dB from the Shannon limits for all considered code rates. In Fig.~\ref{Fig_RequiredSNRMultiRate}, we also include the simulation results of three AR4JA LDPC codes with code rates $1/2$, $2/3$ and $4/5$ in the CCSDS standard~\cite{CCSDS12Coding}, and five PBRL LDPC codes~\cite{Chen15} with code rates $1/4$, $1/3$, $1/2$, $2/3$, and $4/5$, all of which have information length $16384$. We observe that the systematic BMST-R codes have a similar performance as both AR4JA LDPC codes and PBRL LDPC codes over such code rates. Note that no simulation results were reported for AR4JA LDPC codes and PBRL LDPC codes with rates less than $1/4$, while codes of all rates of interest in the interval (0,1) can be constructed using the systematic BMST-R construction. Actually, to the best of our knowledge, no other methods were reported along with simulations in the literature that can construct good rate-compatible codes over such a wide range of code rates.

\end{example}

\subsection{Further Discussions}
All the examples above assume that the subcodewords are modulated using BPSK and transmitted over an AWGN channel. In this subsection, we study the performance of systematic BMST-R codes transmitted over a block fading channel. The word-error-rate~(WER) is defined as the ratio between the number of erroneous subcodewords and the total number of subcodewords transmitted.

Assume that the subcodeword $\boldsymbol{c}^{(t)}$ is modulated using BPSK with 0 and 1 mapped to $+1$ and $-1$, respectively, and transmitted over a block fading channel, resulting in a received vector $\boldsymbol{y}^{(t)}$ expressed as
\begin{equation}\label{TwoChannels}
  y_{j}^{(t)} = a_{j}^{(t)}c_{j}^{(t)} + z_{j}^{(t)}
\end{equation}
for $0 \leq j \leq \Layerk N-\Layerk_p$, where $y_{j}^{(t)}$ is the $j$-th component of $\boldsymbol{y}^{(t)}$, $z_{j}^{(t)}$ is a sample from an independent Gaussian random variable with distribution $\mathcal{N}(0, \sigma^2)$, and $a_{j}^{(t)}$ is a fading coefficient.
In this paper, we consider a Rayleigh fading channel, where $a_{j}^{(t)}$ is a sample from a Rayleigh distribution $\mathcal{R}$ with $\mathbb{E}\left[ \mathcal{R} ^2\right]=1$. For block fading channels with a coherence period of $B_f$ symbols, we assume that $a_{j}^{(t)}$~(perfectly known at the receiver) remains constant over $B_f$ symbols within the same period and is independent identically distributed across different coherence periods.


\begin{figure}[t]
  \centering
  \includegraphics[width=\figwidth]{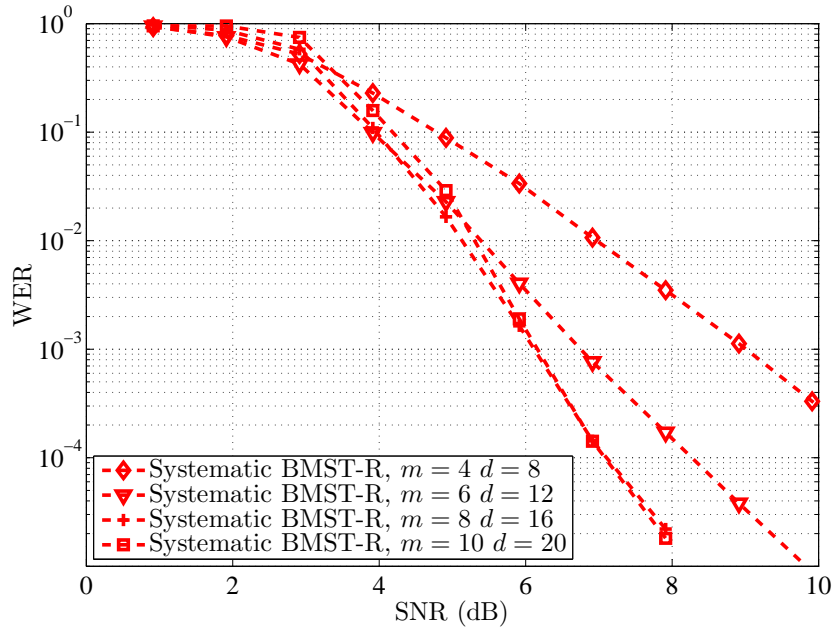}
  \caption{Performance of $R_L=0.49$ systematic BMST-R codes with information subsequence length $\Layerk=100$, repetition degree $N=2$ and puncturing fraction $\theta=0$ over a block fading channel. The encoding memories are $\Memory=4,6,8,$ and $10$. The decoding delay is specified as $d=2\Memory$.}
  \label{Fig_PBMST_Diff_m_FadingChannel}
\end{figure}

\begin{figure}[t]
  \centering
  \includegraphics[width=\figwidth]{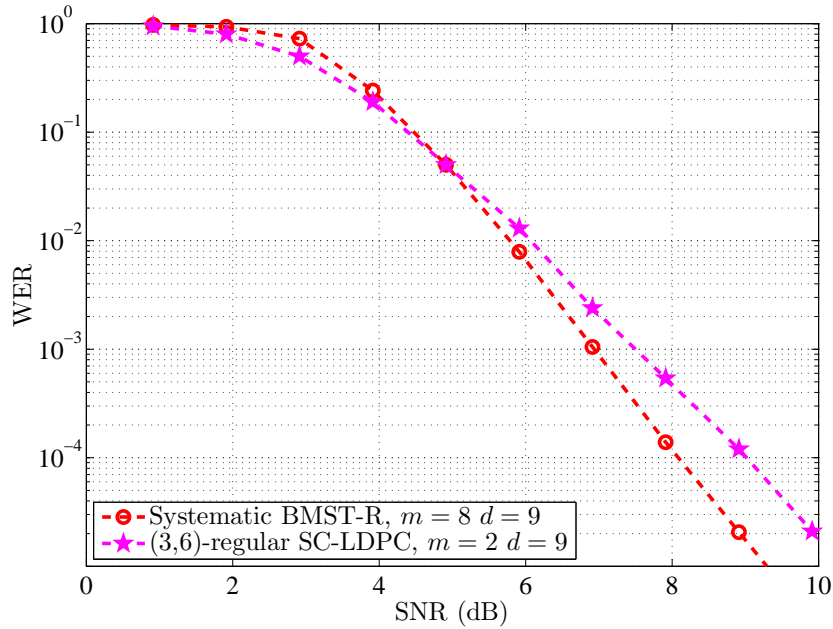}
  \caption{Performance comparison of the systematic BMST-R code and the SC-LDPC code with BPSK modulation over a block fading channel. The systematic BMST-R code is constructed with information subsequence length $\Layerk=100$, encoding memory $\Memory=8$, repetition degree $N=2$, and puncturing fraction $\theta=0$, and decoded with decoding delay $d=9$. The $(3,6)$-regular SC-LDPC codes is constructed with the protograph lifting factor 100 and three component submatrices $\mathbf{B}_{0}=\mathbf{B}_{1}=\mathbf{B}_{2}=[1~1]$, and decoded with decoding delay $d=9$. The decoding latencies of two codes are the same.}
  \label{Fig_PBMST_SCLDPC_FadingChannel}
\end{figure}

\begin{example}
Consider $R_L=0.49$ systematic BMST-R codes with information subsequence length $\Layerk=100$, repetition degree $N=2$ and puncturing fraction $\theta=0$. The subcodewords are modulated using BPSK and transmitted over a block fading channel with a coherence period of $B_f=100$ symbols. That is, a subcodeword $\boldsymbol{c}^{(t)}$ has $F={\Layerk N}/{B_f}=2$ independent fading values. The WER curves of systematic BMST-R codes constructed with encoding memory $\Memory=4,6,8,$ and $10$, and decoded with decoding delay $d=2\Memory$ are shown in Fig.~\ref{Fig_PBMST_Diff_m_FadingChannel}, where we observe that the performance of systematic BMST-R code improves with increasing encoding memory $\Memory$ until $\Memory=8$ and then it degrades slightly as $\Memory$ increases further. This implies that $\Memory=8$ is a good choice for optimum performance.

The performance comparison\footnote{The simulation results~(not included in the figure) suggest that non-systematic BMST codes suffer from severe performance degradation caused by the error propagation.} of the systematic BMST-R code and the SC-LDPC code over a block fading channel is shown in Fig.~\ref{Fig_PBMST_SCLDPC_FadingChannel}. The systematic BMST-R code is constructed with information subsequence length $\Layerk=100$, encoding memory $\Memory=8$, repetition degree $N=2$, and puncturing fraction $\theta=0$, and decoded with decoding delay $d=9$. The $(3,6)$-regular SC-LDPC code is constructed with the protograph lifting factor 100 and three component submatrices $\mathbf{B}_{0}=\mathbf{B}_{1}=\mathbf{B}_{2}=[1~1]$, and decoded with decoding delay $d=9$ presented in~\cite{Hassan14ISIT}. Thus, the decoding latencies of two codes are the same. We see from Fig.~\ref{Fig_PBMST_SCLDPC_FadingChannel} that, in the low WER region, the systematic BMST-R code performs better than the $(3,6)$-regular SC-LDPC code under the equal decoding latency constraint. For example, at a WER of $10^{-4}$, the systematic BMST-R code gains about one dB compared to the equal latency $(3,6)$-regular SC-LDPC code. These results confirm that systematic BMST-R codes without any further optimization can perform well over block fading channels.

\end{example}

\section{Conclusions}\label{sec:Conclusion}
In this paper, we have proposed systematic block Markov superposition transmission~(BMST) of repetition codes, referred to as systematic BMST-R codes. Using both extending and puncturing, systematic BMST-R codes support a wide range of code rates but maintain essentially the same encoding/decoding hardware structure. The systematic BMST-R codes not only preserve the advantages of the original non-systematic BMST codes, namely, low encoding complexity, effective sliding window decoding algorithm and predictable error floors, but also have improved decoding performance especially in short-to-moderate decoding latency. A simple lower bound and an upper bound were derived to analyze the performance of systematic BMST-R codes under MAP decoding. Simulation results show that, over an AWGN channel, 1)~the performance of systematic BMST-R codes around or below the BER of $10^{-5}$ can be predicted by the lower bound; 2)~systematic BMST-R codes can approach the Shannon limit at a BER of $10^{-5}$ within one dB for a wide range of code rates; and 3)~systematic BMST-R codes can outperform both non-systematic BMST codes and SC-LDPC codes in the waterfall region under the equal decoding latency constraint. Simulation results also show that, systematic BMST-R codes without any further optimization can outperform $(3,6)$-regular SC-LDPC codes over a block fading channel. A final note is that the construction of systematic BMST-R codes can be extended to high-order Abelian groups since only addition is required during the encoding process.


\section*{Acknowledgment}
The authors would like to thank Prof. Costello from University of Notre Dame for his helpful comments on the performance lower bounds. The second author, ever visiting University of Notre Dame for one year as an exchange student, is also grateful to Prof. Costello for his insightful supervision on the related research. The authors would also like to thank Dr. Chulong Liang and Dr. Jia Liu for helpful discussions.


\ifCLASSOPTIONcaptionsoff
  \newpage
\fi
\bibliographystyle{IEEEtran}

\end{document}